\documentclass[trackchanges,twocolumn]{aastex701}

\newcommand{\teff}{$T_{\rm eff}$}
\newcommand{\logg}{$\log g$}
\newcommand{\feh}{[Fe/H]}
\newcommand{\vmic}{$\xi$}

\begin{document}

\title{A comprehensive study of the relations between the properties of planetary systems \\ and the chemical compositions of their host stars}

\author[orcid=0000-0002-9089-0136]{Luan Ghezzi}
\email{luanghezzi@ov.ufrj.br}
\correspondingauthor{Luan Ghezzi}
\affiliation{Universidade Federal do Rio de Janeiro, Observatório do Valongo, Ladeira do Pedro Antônio, 43, Rio de Janeiro, RJ 20080-090, Brazil}
\affiliation{Observatório Nacional, Rua General José Cristino, 77, 20921-400 São Cristóvão, Rio de Janeiro, RJ, Brazil}
 
\author[0000-0002-1549-626X]{Ellen Costa-Almeida}
\email{ellenalmeida@on.br}
\affiliation{Observatório Nacional, Rua General José Cristino, 77, 20921-400 São Cristóvão, Rio de Janeiro, RJ, Brazil}

\author[0000-0003-0506-8269]{Verónica Loaiza-Tacuri}
\email{vloatac@academico.ufs.br}
\affiliation{Departamento de Física, Universidade Federal de Sergipe, Av. Marcelo Deda Chagas, s/n, 49107-230, São Cristóvão, SE, Brazil}

\author[0000-0001-6476-0576]{Katia Cunha}
\email{kcunha@arizona.edu}
\affiliation{Steward Observatory, University of Arizona, 933 North Cherry Avenue, Tucson, AZ 85721, USA}
\affiliation{Observatório Nacional, Rua General José Cristino, 77, 20921-400 São Cristóvão, Rio de Janeiro, RJ, Brazil}


\begin{abstract}

The giant planet--metallicity correlation revealed that planetary formation depends on the stellar properties. There is growing evidence that it is also valid for smaller hot planets, but it is not clear whether elements other than iron also influence the properties of planetary systems. To investigate this, we determined the abundances of 13 chemical elements (Na, Mg, Al, Si, Ca, Sc, Ti, V, Cr, Mn, Co, Ni, and Cu) for a sample of 561 Kepler exoplanet-hosting stars using high-resolution Keck/HIRES spectra. We find that stars in systems having only large or hot planets are enriched in some elements relative to those having only small or warm planets, respectively, with this signature being related to the underlying stellar metallicity. This Kepler sample is composed of stars belonging to the Galactic low- and high-$\alpha$ sequences, corresponding to the chemical thin and thick disks. Our results reveal that stars enhanced in $\alpha$-elements may facilitate the formation of large planets in metal-poor environments although the iron abundance is still a limiting factor. We also investigated chemical abundances as a function of elemental condensation temperatures and found that there is a diversity of slopes regardless of the exoplanetary systems hosted by the star. We confirmed that the Sun is depleted in refractory elements relative to the solar twins in our sample, all of which host a diversity of exoplanets, suggesting that this depletion is caused by processes not related to planet formation.  

\end{abstract}



\section{Introduction} 
\label{sec:intro}

The formation, evolution, architectures, and habitability of planetary systems depend on the physical properties and chemical compositions of their host stars. The strongest evidence for this connection is the planet-metallicity correlation, first proposed by \cite{gonzalez97} and subsequently confirmed by multiple studies \citep[e.g.,][]{santos04,fischer05,ghezzi10,ghezzi18}. Independent analyses of different samples consistently showed that the frequency of giant planets increases with the stellar metallicity for FGK main-sequence stars. This relation provides strong support for the core accretion mechanism of planet formation \citep{il04}.

Follow-up studies investigated whether this correlation is also valid for smaller planets. Due to technical limitations at the time, these were based on small samples but revealed a different picture. Stars hosting only Neptunian-mass planets were not preferentially metal rich \citep[e.g.,][]{sousa08,ghezzi10}. Leveraging the much larger sample of stars with small planets discovered by the Kepler mission \citep{borucki10}, \cite{buchhave12} showed that planets with $R_{pl} < 4 R_{\oplus}$ orbit stars with a wide range of metallicities, while giant planets are preferentially found around more metal-rich hosts \citep[but see also][]{wf15}. In a subsequent study, \cite{buchhave14} classified 600 exoplanet candidates into three different groups according to the metallicity distributions of their $\sim$400 parent stars, suggesting that it is an important parameter in shaping the architectures of planetary systems. 

Using a sample of more than 20,000 Kepler stars that included 665 planet candidates, \cite{mulders16} further refined this result showing that the occurrence rates of hot exoplanets ($P \leq$ 10 days) were almost 3 times higher for stars that are more metal rich than the Sun compared to their more metal-poor counterparts. \cite{wilson18} found a similar separation from the analysis of 282 Kepler Objects of Interest (KOIs) observed by the Sloan Digital Sky Survey (SDSS) IV Apache Point Observatory Galactic Evolution Experiment \citep[APOGEE; ][]{majewski17}, although for a slightly smaller orbital period ($P \sim$ 8 days). 

Analyzing an even larger sample of 1305 planet-hosting stars from the California-Kepler Survey \citep[CKS; ][]{petigura17}, \cite{petigura18} concluded that the occurrence rate of exoplanets depends on stellar metallicity, but with varying strengths according to the planetary radii and orbital periods. For instance, the occurrence of super-Earths ($R_{pl} = 1.0 - 1.7 R_{\oplus}$) increases with stellar metallicities for the hot planets ($P \leq$ 10 days) but not for their warm counterparts (10 days $ < P \leq$ 100 days). Also using a sample taken from the CKS, \cite{ghezzi21} found a statistically significant difference between the metallicity distributions of systems with single hot super-Earths and single warm super-Earths. However, a similar result was not observed for systems with multiple super-Earths or with single or multiple sub-Neptunes ($R_{pl} = 1.9 - 4.4 R_{\oplus}$). These efforts revealed that the planet-metallicity correlation is much more complex than initially thought \citep[see also][]{adibekyan19,teske24}.

Complementary studies investigated whether elements other than iron might also be related to the process of planetary formation. For instance, \cite{robinson06} found that stars hosting giant planets were more enriched in silicon and nickel relative to a population of metal-rich stars without known planets. \cite{brugamyer11} provided additional support for the dependence of the occurrence rate of giant planets on stellar silicon abundances, however they did not find a similar result for oxygen. \cite{adibekyan12a,adibekyan12b} presented evidence that planet occurrence is higher among stars with larger abundances of $\alpha$-elements in the metal-poor regime, suggesting that other elements are important for planetary formation when small amounts of iron are available. 

More recently, \cite{wilson22} analyzed a sample of 1018 KOIs that also had chemical abundances for 10 elements (C, Mg, Al, Si, S, K, Ca, Mn, Fe, and Ni) from APOGEE \citep{majewski17}. They found that the occurrence rates of hot super-Earths ($R_{pl} = 1.0 - 1.9 R_{\oplus}$) and sub-Neptunes ($R_{pl} = 1.9 - 4.0 R_{\oplus}$) increase for higher abundances of any of the elements, with a stronger correlation for the latter class. These results extended those of \cite{petigura18} for nine additional elements. 

Instead of focusing on individual elements, \cite{torres-quijano25} applied a machine learning algorithm to an updated version of the Hypatia Catalog \citep{hinkel14} to search for abundance patterns that could indicate the presence of small planets ($R_{pl} = 1.0 - 3.5 R_{\oplus}$). Surprisingly, different experiments revealed that sodium (Na) and vanadium (V) were always among the most important elements and that aluminum (Al) also seemed to be relevant \citep[in agreement with the results of][for Al]{wilson22}.

Current evidence from multiple studies thus suggests that the chemical abundances of different elements influence the formation of planets. However, this process might also affect the chemical composition of the host stars. \cite{melendez09} showed that the Sun is depleted in refractory elements (i.e., with condensation temperatures $T_{C} <$ 900 K) relative to the solar twins. They suggested that this could be a signature of the formation of rocky planets, since the ``missing'' material would be locked up in the terrestrial planets. \cite{schuler15} investigated this hypothesis using a sample of seven stars with at least one small planet ($R_{pl} \leq 1.6 R_{\oplus}$) and did not find the signature for any of them. Subsequent studies based on larger samples revealed that, although the Sun is indeed depleted in refractory elements compared to solar analogs or twins, this difference does not seem to be related to planetary formation, since stars with planets exhibit a diversity of trends of elemental abundances as a function of condensation temperature (e.g., \citealt{bedell18,liu20,nibauer21,rampalli24,martos25,carlos25,sun25a,sun25b,rampalli25}; see also the earlier study from \citealt{smith01} and the recent review by \citealt{gustafsson25}). However, the samples of planet-hosting stars analyzed were relatively small ($\leq$50), while the sample of stellar hosts analyzed in this study is much larger and can offer further insights into this question.

In this paper, we build upon these previous works and that of \cite{ghezzi21} to further investigate possible correlations between the formation and architectures of planetary systems and detailed chemical abundances of their host stars. We use a subsample from the CKS since its large size ($>$ 500 planet-hosting stars) allows statistical conclusions to be drawn for a variety of planetary systems. We provide independent determinations of the abundances of 13 elements (Na, Mg, Al, Si, Ca, Sc, Ti, V, Cr, Mn, Co, Ni, Cu), three of which (Sc, Co, and Cu) were never analyzed before for this sample. This paper is organized as follows. In Section \ref{sec:analysis}, we describe our sample, the data used, and the determination of chemical abundances and stellar evolutionary parameters. We present our results, perform consistency checks, and compare them with the literature in Section \ref{sec:results}. We discuss our findings in Section \ref{sec:discussion} and present our concluding remarks in Section \ref{sec:conclusions}.

\section{Analysis} 
\label{sec:analysis}

\subsection{Sample and Data} 
\label{sec:data}

\begin{figure}
\epsscale{1.0}
\plotone{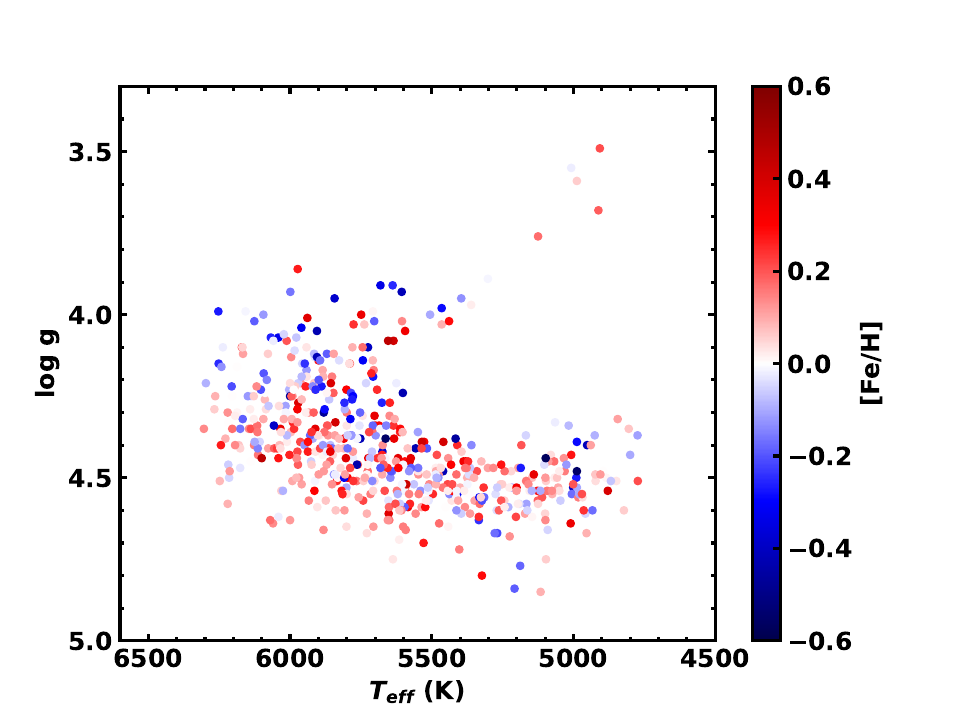}
\caption{Kiel diagram for our sample of 561 stars. The stellar metallicities are represented by the color bar. 
\label{sample}}
\end{figure}

We selected our sample from \cite{ghezzi21}, which originally consisted of 663 planet-hosting stars. This was the ``clean'' sample obtained by \cite{martinez19} from the CKS catalog \citep{petigura17}. However, \cite{ghezzi21} removed 102 stars that had large ($>12\%$) median uncertainties on the equivalent widths (EWs) measured for the iron lines and/or microturbulence velocities greater than 1.7 km s$^{-1}$. These quality cuts provided a final sample of 561 planet-hosting stars that will be analyzed in this study. Their distances, retrieved from \cite{bj21}, range from $\sim$60 to 1800 pc, with a median value of $\sim$630 pc. For this sample, we adopt the precise and homogeneous stellar parameters derived in \cite{martinez19} and \cite{ghezzi21}, shown in Table \ref{table_param} and the Kiel diagram presented in Figure \ref{sample}.

\begin{splitdeluxetable*}{lrrrrrrrrrBrrrrrrrrrrBrrrrrrrrr}
\tablecaption{Parameters determined for the stars in our sample. The columns are: (1) name of the star, (2-4) atmospheric parameters, (5-21) chemical abundances, (22-24) evolutionary parameters and (25-29) slopes and corresponding statistics.
\label{table_param}
}
\tabletypesize{\scriptsize}
\tablehead{\colhead{Star} & \colhead{T$_{eff}$} & \colhead{log g} & \colhead{V$_{mic}$} & \colhead{[Na/H]} & \colhead{[Mg/H]} & \colhead{[Al/H]} & \colhead{[Si/H]} & \colhead{[Ca/H]} & \colhead{[ScI/H]} & \colhead{[ScII/H]} & \colhead{[TiI/H]} & \colhead{[TiII/H]} & \colhead{[V/H]} & \colhead{[CrI/H]} & \colhead{[CrII/H]} & \colhead{[Mn/H]} & \colhead{[Fe/H]} & \colhead{[Co/H]} & \colhead{[Ni/H]} & \colhead{[Cu/H]} & \colhead{M} & \colhead{R} & \colhead{Age} & \colhead{Slope} & \colhead{$t$-value} & \colhead{$p$-value} & \colhead{$p$-value} & \colhead{$p$-value} \\
\colhead{} & \colhead{(K)} & \colhead{(dex)} & \colhead{(km s$^{-1}$)} & \colhead{(dex)} & \colhead{(dex)} & \colhead{(dex)} & \colhead{(dex)} & \colhead{(dex)} & \colhead{(dex)} & \colhead{(dex)} & \colhead{(dex)} & \colhead{(dex)} & \colhead{(dex)} & \colhead{(dex)} & \colhead{(dex)} & \colhead{(dex)} & \colhead{(dex)} & \colhead{(dex)} & \colhead{(dex)} & \colhead{(dex)} & \colhead{(M$_{\odot}$)} & \colhead{(R$_{\odot}$)} & \colhead{(Gyr)} & \colhead{(10$^{-5}$ dex K$^{-1}$)} & \colhead{(Slope)} & \colhead{(Slope)} & \colhead{(Spearman)} & \colhead{(Pearson)}
}
\startdata
Sun (Vesta) & 5792$\pm$16 & 4.47$\pm$0.06 & 1.014$\pm$0.030 & -0.054$\pm$0.009 & -0.015$\pm$0.020 & 0.012$\pm$0.013 & 0.028$\pm$0.007 & 0.028$\pm$0.018 & 0.009$\pm$0.032 & 0.029$\pm$0.026 & 0.027$\pm$0.017 & 0.047$\pm$0.026 & 0.017$\pm$0.019 & 0.013$\pm$0.016 & 0.062$\pm$0.028 & 0.002$\pm$0.016 & 0.020$\pm$0.010 & 0.024$\pm$0.017 & 0.021$\pm$0.011 & -0.005$\pm$0.019 & \nodata & \nodata & \nodata & \nodata & \nodata & \nodata & \nodata & \nodata \\
k00007 & 5852$\pm$20 & 4.19$\pm$0.06 & 1.200$\pm$0.030 & 0.072$\pm$0.051 & 0.172$\pm$0.025 & 0.228$\pm$0.013 & 0.171$\pm$0.010 & 0.185$\pm$0.020 & 0.204$\pm$0.019 & 0.342$\pm$0.036 & 0.187$\pm$0.021 & 0.235$\pm$0.029 & 0.181$\pm$0.022 & 0.146$\pm$0.019 & 0.190$\pm$0.031 & 0.133$\pm$0.018 & 0.150$\pm$0.010 & 0.162$\pm$0.018 & 0.187$\pm$0.014 & 0.172$\pm$0.088 & 1.207$\pm$0.048 & 1.651$\pm$0.023 & 5.066$\pm$0.780 & 22$\pm$5 & 4.104 & $1.746 \times 10^{-3}$ & $7.200 \times 10^{-5}$ & $9.701 \times 10^{-4}$ \\
k00017 & 5699$\pm$22 & 4.34$\pm$0.07 & 1.016$\pm$0.040 & 0.527$\pm$0.022 & 0.374$\pm$0.045 & 0.370$\pm$0.012 & 0.402$\pm$0.019 & 0.315$\pm$0.024 & 0.418$\pm$0.028 & 0.453$\pm$0.033 & 0.402$\pm$0.026 & 0.414$\pm$0.032 & 0.400$\pm$0.026 & 0.371$\pm$0.023 & 0.353$\pm$0.036 & 0.407$\pm$0.023 & 0.360$\pm$0.010 & 0.430$\pm$0.019 & 0.407$\pm$0.014 & 0.547$\pm$0.026 & 1.136$\pm$0.018 & 1.323$\pm$0.016 & 5.348$\pm$0.600 & -17$\pm$8 & -2.226 & $4.790 \times 10^{-2}$ & $2.043 \times 10^{-1}$ & $2.352 \times 10^{-2}$ \\
k00020 & 6089$\pm$28 & 4.26$\pm$0.04 & 1.256$\pm$0.040 & -0.029$\pm$0.041 & 0.100$\pm$0.037 & 0.147$\pm$0.025 & 0.067$\pm$0.011 & 0.125$\pm$0.022 & 0.146$\pm$0.030 & 0.177$\pm$0.020 & 0.118$\pm$0.027 & 0.186$\pm$0.020 & 0.118$\pm$0.029 & 0.065$\pm$0.025 & 0.086$\pm$0.024 & 0.004$\pm$0.024 & 0.080$\pm$0.010 & 0.117$\pm$0.028 & 0.087$\pm$0.019 & 0.084$\pm$0.049 & 1.213$\pm$0.036 & 1.515$\pm$0.018 & 4.052$\pm$0.551 & 24$\pm$4 & 5.271 & $2.640 \times 10^{-4}$ & $5.000 \times 10^{-6}$ & $9.123 \times 10^{-5}$ \\
k00022 & 5964$\pm$25 & 4.39$\pm$0.11 & 1.147$\pm$0.040 & 0.278$\pm$0.022 & 0.209$\pm$0.022 & 0.258$\pm$0.026 & 0.237$\pm$0.010 & 0.233$\pm$0.026 & 0.271$\pm$0.042 & 0.336$\pm$0.046 & 0.206$\pm$0.027 & 0.275$\pm$0.045 & 0.297$\pm$0.028 & 0.239$\pm$0.025 & 0.234$\pm$0.042 & 0.230$\pm$0.021 & 0.220$\pm$0.020 & 0.278$\pm$0.026 & 0.273$\pm$0.017 & 0.194$\pm$0.092 & 1.173$\pm$0.010 & 1.264$\pm$0.018 & 3.536$\pm$0.373 & 12$\pm$4 & 2.843 & $1.600 \times 10^{-2}$ & $1.491 \times 10^{-1}$ & $1.083 \times 10^{-1}$ \\
k00041 & 5915$\pm$15 & 4.21$\pm$0.04 & 1.275$\pm$0.030 & 0.090$\pm$0.049 & 0.089$\pm$0.014 & 0.093$\pm$0.020 & 0.126$\pm$0.006 & 0.051$\pm$0.017 & 0.092$\pm$0.019 & 0.156$\pm$0.025 & 0.089$\pm$0.016 & 0.120$\pm$0.020 & 0.089$\pm$0.016 & 0.050$\pm$0.015 & 0.122$\pm$0.025 & 0.048$\pm$0.019 & 0.070$\pm$0.010 & 0.103$\pm$0.015 & 0.097$\pm$0.010 & 0.138$\pm$0.019 & 1.139$\pm$0.014 & 1.585$\pm$0.001 & 6.101$\pm$0.230 & 3$\pm$4 & 0.725 & $4.835 \times 10^{-1}$ & $4.643 \times 10^{-1}$ & $6.211 \times 10^{-1}$ \\
k00046 & 5635$\pm$26 & 4.08$\pm$0.06 & 1.129$\pm$0.040 & 0.288$\pm$0.026 & 0.519$\pm$0.031 & 0.500$\pm$0.022 & 0.441$\pm$0.013 & 0.413$\pm$0.031 & 0.404$\pm$0.073 & 0.573$\pm$0.045 & 0.473$\pm$0.034 & 0.549$\pm$0.037 & 0.470$\pm$0.031 & 0.454$\pm$0.034 & 0.456$\pm$0.058 & 0.356$\pm$0.031 & 0.410$\pm$0.020 & 0.456$\pm$0.022 & 0.412$\pm$0.018 & 0.415$\pm$0.031 & 1.293$\pm$0.043 & 1.855$\pm$0.042 & 4.412$\pm$0.482 & 30$\pm$7 & 4.473 & $9.430 \times 10^{-4}$ & $1.123 \times 10^{-2}$ & $1.188 \times 10^{-3}$ \\
k00049 & 5858$\pm$41 & 4.40$\pm$0.12 & 1.329$\pm$0.090 & -0.076$\pm$0.049 & -0.052$\pm$0.054 & \nodata & -0.044$\pm$0.028 & -0.072$\pm$0.035 & 0.166$\pm$0.232 & 0.013$\pm$0.050 & 0.008$\pm$0.045 & 0.043$\pm$0.056 & 0.047$\pm$0.065 & -0.114$\pm$0.048 & -0.056$\pm$0.059 & -0.223$\pm$0.044 & -0.080$\pm$0.030 & -0.147$\pm$0.053 & -0.144$\pm$0.032 & -0.043$\pm$0.054 & 1.024$\pm$0.016 & 1.417$\pm$0.033 & 8.324$\pm$0.515 & 19$\pm$11 & 1.709 & $1.183 \times 10^{-1}$ & $1.446 \times 10^{-1}$ & $1.313 \times 10^{-1}$ \\
k00063 & 5594$\pm$22 & 4.53$\pm$0.06 & 1.229$\pm$0.040 & 0.036$\pm$0.040 & 0.130$\pm$0.033 & 0.160$\pm$0.012 & 0.143$\pm$0.012 & 0.209$\pm$0.029 & 0.179$\pm$0.039 & 0.198$\pm$0.032 & 0.208$\pm$0.026 & 0.169$\pm$0.031 & 0.192$\pm$0.026 & 0.187$\pm$0.026 & 0.163$\pm$0.043 & 0.127$\pm$0.034 & 0.160$\pm$0.010 & 0.115$\pm$0.029 & 0.099$\pm$0.013 & -0.028$\pm$0.052 & 1.017$\pm$0.015 & 0.915$\pm$0.011 & 1.315$\pm$1.004 & 29$\pm$6 & 5.108 & $3.398 \times 10^{-4}$ & $1.317 \times 10^{-2}$ & $1.247 \times 10^{-3}$ \\
k00069 & 5646$\pm$21 & 4.50$\pm$0.04 & 0.908$\pm$0.030 & -0.160$\pm$0.013 & -0.108$\pm$0.020 & -0.100$\pm$0.017 & -0.101$\pm$0.007 & -0.114$\pm$0.022 & -0.139$\pm$0.040 & -0.062$\pm$0.027 & -0.104$\pm$0.022 & -0.061$\pm$0.021 & -0.114$\pm$0.023 & -0.150$\pm$0.020 & -0.068$\pm$0.023 & -0.209$\pm$0.020 & -0.140$\pm$0.010 & -0.135$\pm$0.018 & -0.148$\pm$0.012 & -0.161$\pm$0.036 & 0.883$\pm$0.011 & 0.989$\pm$0.019 & 10.544$\pm$0.820 & 16$\pm$3 & 4.694 & $6.563 \times 10^{-4}$ & $7.570 \times 10^{-4}$ & $1.405 \times 10^{-3}$ \\
\nodata & \nodata & \nodata & \nodata & \nodata & \nodata & \nodata & \nodata & \nodata & \nodata & \nodata & \nodata & \nodata & \nodata & \nodata & \nodata \\
\enddata
\tablecomments{This table is published in its entirety in the machine-readable format. A portion is shown here for guidance regarding its form and content.}
\end{splitdeluxetable*}

The planet radii for the sample were calculated by \cite{martinez19}, and the corresponding planetary systems were classified by \cite{ghezzi21} according to all confirmed planets orbiting the host star. Briefly, planets with $R_{pl} < 1.9 R_{\oplus}$, $1.9 R_{\oplus} \leq R_{pl} < 4.4 R_{\oplus}$, $4.4 R_{\oplus} \leq R_{pl} < 8.0 R_{\oplus}$ and $R_{pl} > 8.0 R_{\oplus}$ are considered super-Earths (SE), sub-Neptunes (SN), sub-Saturns (SS) and Jupiters (JP), respectively. Planets with orbital periods $P \leq$ 10 days and 10 days $ < P \leq 100$ days are classified as hot (H) and warm (W), respectively. Finally, systems with only one or more than one detected planet are considered, respectively, single or multiple. For instance, a system classified as ``SE single H'' has one hot super-Earth while one in the category ``SN multi W'' has at least two warm sub-Neptunes.

The spectra are the same as analyzed by \cite{ghezzi21} and were obtained by the CKS team \citep{petigura17} with the High Resolution Echelle Spectrometer \citep[HIRES;][]{vogt94} spectrograph attached to the Keck I 10 m telescope (Maunakea, Hawaii). The spectra have high resolution (R $\sim$ 60,000), almost complete wavelength coverage between 3640 \AA\ and 7990 \AA, and are publicly available\footnote{\url{https://california-planet-search.github.io/cks-website/}}. Their signal-to-noise ratios (S/Ns) vary from $\sim$20 to $\sim$300 and have a median value of $\sim$60, as measured by \cite{ghezzi21}.

\subsection{Line List} 
\label{sec:line_list}

We compiled an initial list of 231 lines for the 16 species analyzed in this work (Na I, Mg I, Al I, Si I, Ca I, Sc I, Sc II, Ti I, Ti II, V, Cr I, Cr II, Mn I, Co I, Ni I, Cu I) from the following sources: \cite{melendez14}, \cite{bedell14}, \cite{schuler15}, \cite{teske15}, \cite{bedell18} and \cite{liu18}. Their atomic parameters were retrieved from the Vienna Atomic Line Database \citep{piskunov95,ryabchikova97,kupka99,kupka00,ryabchikova15,pakhomov19}. For Sc, V, Mn, Co, and Cu, we considered the hyperfine structure and isotopic splittings. We visually inspected all lines in the Solar Atlas spectrum from \cite{hinkle00}, which is the one originally published by \cite{kurucz84} but corrected for telluric absorption \citep{melendez06}, and removed eight lines that were too weak, blended, or close to strong lines (e.g., H$\alpha$).

We automatically measured the equivalent widths (EWs) for the remaining lines using ARES v2 \citep{sousa07,sousa15} with the following input parameters: \textit{smoothder} = 4, \textit{space} = 3.0, \textit{rejt} = 0.999, \textit{lineresol} = 0.1, and \textit{miniline} = 5. We noted that ARES did not provide good fits for some lines, so we manually measured their EWs using the \texttt{splot} task in the Community Distribution of IRAF \citep{tody86,tody93,iraf_ascl,iraf_zenodo}\footnote{Image Reduction and Analysis Facility (IRAF) was written by the National Optical Astronomy Observatories (NOAO) and is now maintained by the iraf-community. IRAF is listed in the Astronomical Source Code Library as ascl:9911.002. The DOI is 10.5281/zenodo.5816743.}.

Using this revised line list and the 2019 version of the software MOOG \citep{sneden73}\footnote{\url{https://www.as.utexas.edu/~chris/moog.html}}, we determined the chemical abundances for the Sun. The model atmosphere was adopted from the KURUCZ ATLAS9 ODFNEW grid \citep{ck03} and has the following parameters: \teff = 5777 K, \logg\ = 4.44, \feh\ = 0.00 and \vmic\ = 1.00 km s$^{-1}$. We used the driver \textit{blends} on MOOG for Sc, V, Mn, Co, and Cu and \textit{abfind} for the other elements, both with option 1 for the treatment of damping, i.e., van der Waals damping factors within the code \citep{barklem00,barklem05} are adopted if available; otherwise, values from our line list are utilized. After checking the results, we adjusted the $\log$ \textit{gf} values so that all lines for a given element returned the corresponding solar abundances from \cite{asplund09}. Finally, we removed 18 lines that are located in the interorder or inter-CCD spaces in the HIRES spectra. Our final list contains 205 lines and can be seen in Table \ref{table_line_list_ews}, which also contains the measured EWs for the Solar Atlas.

\begin{deluxetable}{llrrrrr}[ht]
\tablecaption{Line list 
\label{table_line_list_ews}
}
\tablehead{
\colhead{Star} & \colhead{$\lambda$} & \colhead{ID\tablenotemark{a}} & \colhead{$\chi$} & \colhead{$\log gf$} & \colhead{$\log \Gamma_{w}$\tablenotemark{b}} & \colhead{EW} \\
\colhead{} & \colhead{(\AA)} & \colhead{} & \colhead{(eV)} & \colhead{} &
\colhead{($sN_{H}^{-1}$)} & \colhead{(m\AA)}
}
\startdata
Sun & 4751.822 & 11.0 & 2.104 & -2.078 &  0.000 &  13.0 \\
Sun & 5148.838 & 11.0 & 2.102 & -2.044 &  0.000 &  12.5 \\
Sun & 6154.225 & 11.0 & 2.102 & -1.547 &  0.000 &  37.0 \\
Sun & 6160.747 & 11.0 & 2.104 & -1.246 &  0.000 &  56.2 \\
Sun & 4571.096 & 12.0 & 0.000 & -5.623 & -7.770 & 108.7 \\
Sun & 4730.029 & 12.0 & 4.346 & -2.347 &  0.000 &  69.8 \\
Sun & 5711.090 & 12.0 & 4.346 & -1.724 &  0.000 & 104.7 \\
Sun & 6318.717 & 12.0 & 5.108 & -2.103 &  0.000 &  38.5 \\
Sun & 6319.237 & 12.0 & 5.108 & -2.324 &  0.000 &  26.2 \\
Sun & 5557.063 & 13.0 & 3.143 & -2.110 &  0.000 &  12.4 \\
\nodata & \nodata & \nodata & \nodata & \nodata & \nodata \\
\enddata
\tablecomments{This table is published in its entirety in the machine-readable format. A portion is shown here for guidance regarding its form and content.}
\tablenotetext{a}{Following the MOOG format, each species is represented by its atomic number followed by a number after the decimal place: 0 for neutral and 1 for singly ionized species.}
\tablenotetext{b}{$\log \Gamma_{w}$ is the logarithm of the van der Waals damping constant at 10,000 K.}
\end{deluxetable}

\subsection{Chemical Abundances} 
\label{sec:abundances}

The chemical abundances for the stars in our sample were obtained in a similar manner as for the Solar Atlas. The model atmospheres were interpolated from the KURUCZ ATLAS9 ODFNEW grid \citep{ck03} using the atmospheric parameters (\teff, \logg, \feh, and \vmic) from \cite{martinez19} and \cite{ghezzi21}. The EWs were automatically measured with ARES v2 \citep{sousa07,sousa15} with the same input parameters described in Section \ref{sec:line_list}, except for the \textit{rejt} parameter, which was replaced by the S/N value of the spectrum of each star. 

Note, however, that EWs could not be measured for all lines because some of them were too weak or blended, depending on the stellar parameters and S/N of the spectra. In particular, ARES was not able to measure any EWs for Sc I for seven stars: KOI-244, KOI-262, KOI-1530, KOI-2833, KOI-3165, KOI-3438, and KOI-3928. They are hotter stars with lower metallicities, so Sc I lines are very weak (EW $<$ 5 m\AA).

For each star, we removed lines with measured depths lower than zero or higher than one, as these are nonphysical values. Additionally, we cut lines with relative errors on the EWs larger than 50\%, which indicate poor fits to the profiles, or with EW $>$ 150 m\AA, to avoid very strong lines. The cut based on the relative errors left no Al lines for the stars KOI-49, KOI-620, KOI-1779, and KOI-2022. All EWs used for the following abundance determination are provided in Table \ref{table_line_list_ews}.

We determined the chemical abundances using the driver \textit{blends} of the 2019 version of MOOG \citep{sneden73} for elements that have hyperfine structure and isotopic splittings (Sc, V, Mn, Co, and Cu) and \textit{abfind} otherwise, both with option 1 for the damping input parameter. For species that had two or more lines, we made an initial cut of lines having abundances with absolute deviations from the median larger than 0.40 dex to remove lines with possible problems in their EW measurements. This procedure left no Na lines for the star KOI-837. Then, we performed a cut of lines for which abundances were outside a two-median absolute deviation (MAD) interval around the median. However, we noted that this procedure was not enough to remove bad measurements in all cases, so we added two additional clippings to remove abundances with absolute deviations greater than 0.20 dex. This iterative process ensured that we ended up with precise abundances even for stars with the lowest S/N spectra. Finally, we subtracted the mean abundances from the reference solar values given by \cite{asplund09} to obtain [X/H] values, where X is a given element. The abundances for all stars in our sample are shown in Table \ref{table_param}, which also includes the values obtained from the analysis of the spectrum of sunlight reflected off Vesta (see Section \ref{sec:sun}).

We calculated the abundance uncertainties considering the contributions from the dispersions in the abundances returned by the individual lines as well as the variations caused by the uncertainties in each of the four atmospheric parameters (\teff, \logg, \feh, and \vmic). The former was estimated from the standard deviation of the mean for the absolute abundance A(X), if the element X had abundance measurements from two or more lines. The contribution of the effective temperature for the uncertainty was obtained by generating two new model atmospheres, with \teff $\pm \sigma$(\teff), and determining two new sets of abundances for all elements. Then, we calculated the differences A(X)$_{final}$ - A(X)$_{\rm T_{eff} \pm \sigma(\rm T_{eff})}$, and the maximum absolute value was taken as the uncertainty caused by the error on \teff. A similar procedure was done for estimating the uncertainties caused by the errors in \logg, \feh, and \vmic. The total uncertainties for the abundances of each element were determined by adding all five contributions in quadrature and are shown in Table \ref{table_param}. 

The largest median uncertainty is 0.053 $\pm$ 0.012 for Cr II, while Si has the most precise abundances with a median error of 0.019 $\pm$ 0.005. These results show that our methodology produced reliable abundances given the low S/N values of most of the spectra, although we note that the dispersions of the errors become greater for S/N $<$ 100 and there is a notable increase of the typical values for S/N $<$ 50 for all species.

\subsection{Stellar Evolutionary Parameters} 
\label{sec:evol_par}

We determined the evolutionary parameters for our stars using the version 1.3 of PARAM \citep{dasilva06}\footnote{\url{https://stev.oapd.inaf.it/cgi-bin/param\_1.3}}. The code employs a Bayesian statistical framework to estimate masses, radii, and ages through a comparison of observational parameters with a grid of isochrones. We chose version 1.1 of the grid of PARSEC isochrones \citep{bressan12} and kept the default options for the Bayesian priors. We also provided the following input parameters: \teff, \feh, parallax, and V magnitude. The effective temperatures were taken from \cite{martinez19}, and the metallicities were from \cite{ghezzi21}. The parallaxes were retrieved from Gaia Data Release 3 \citep[DR3;][]{gaia23}, but two stars (KOI-4098 and KOI-4580) did not have available values. The V magnitudes were calculated from the G magnitudes and $G_{BP}-G_{RP}$ colors following E. Costa-Almeida et al. (2026, in preparation). These magnitudes were then corrected for extinction using $A_{V}$ values calculated with the \texttt{StarHorse} code \citep{queiroz18,anders19,anders22} and available in the contributions to Gaia Early Data Release 3 \citep[EDR3;][]{gaia21} for 549 stars and Gaia Data Release 2 \citep[DR2;][]{gaia18} for 9 stars\footnote{\url{https://gaia.aip.de/cms/data/contrib/}}. For one star, $A_{V}$ was taken from the Galactic Dust Reddening and Extinction tool on the NASA/IPAC Infrared Science Archive (IRSA)\footnote{\url{https://irsa.ipac.caltech.edu/applications/DUST/}}, adopting the value calculated with the extinction map of \cite{schlafly11}. PARAM was unable to find a solution for eight stars. The masses, radii, and ages for the remaining 551 stars are shown in Table \ref{table_param}.

\section{Results} 
\label{sec:results}

\subsection{Chemical abundance distributions}
\label{sec:dist_ab}

\begin{figure*}
\epsscale{1.0}
\plotone{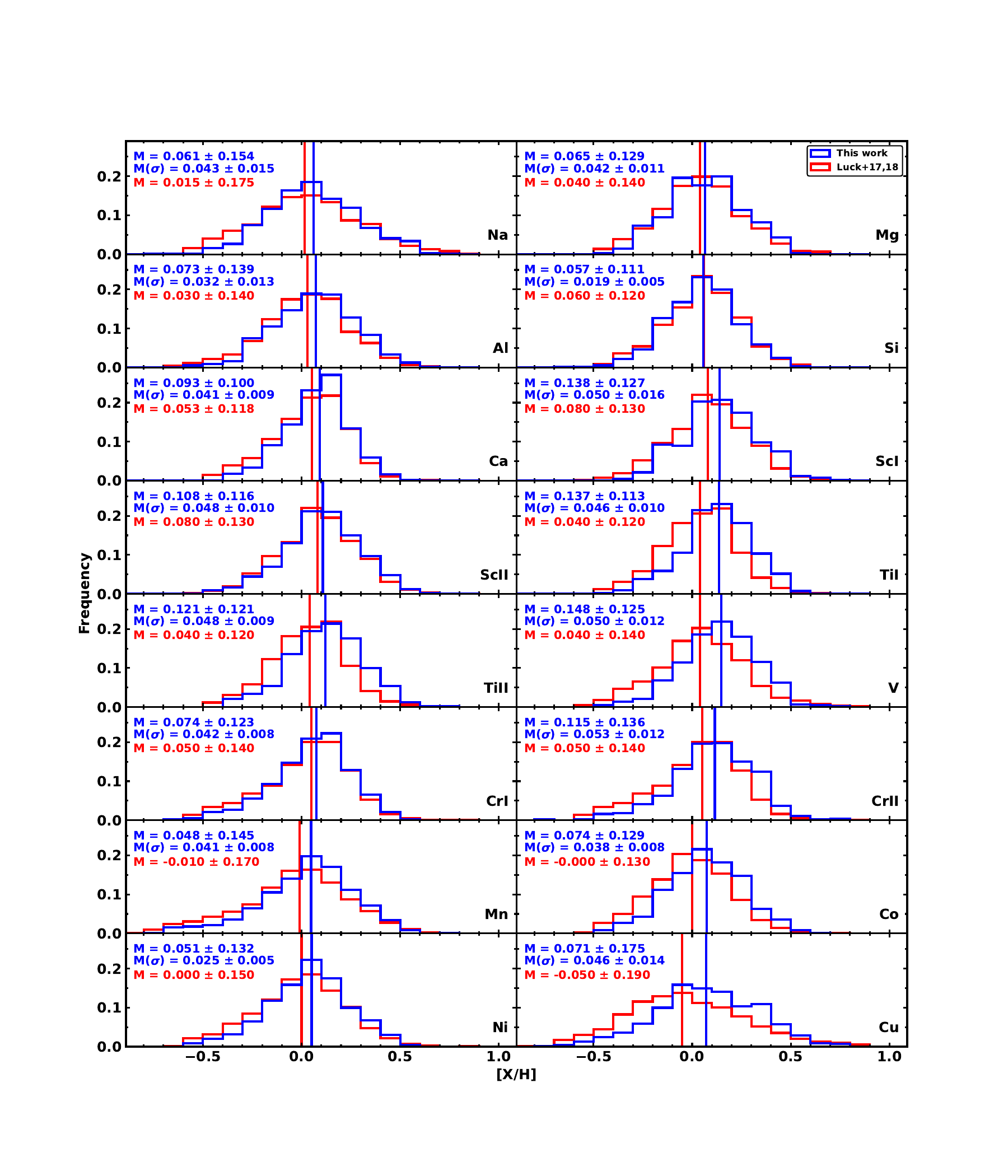}
\caption{Distributions of 16 chemical abundances (13 elements, including three - Sc, Ti and Cr - with two ionization stages) for the 561 stars in our work (blue) and 1368 stars from \cite{luck17,luck18} (red). Note that they provide a single set of abundances for Sc, Ti, and Cr, and we use them in the comparisons with both ionization stages we analyzed for these elements. The solid vertical blue and red lines represent the median values (M), which are also shown in each panel along with the corresponding MAD, from our work and \cite{luck17,luck18}, respectively. The median M($\sigma$) and MAD values for the total uncertainties in our abundances are also shown in each panel.
\label{ab_hist}}
\end{figure*}

The distributions of the chemical abundances determined in this work are shown as blue histograms in Figure \ref{ab_hist}, and their median values are shown as blue solid lines. We can see that our sample, which is composed only of planet hosts, has higher median abundances than the Sun for all elements. This is probably a consequence of the higher metallicity of our sample, having a median and MAD of 0.06 $\pm$ 0.11 dex. As the distributions for all species but Cr I, Cr II, and Mn are non-Gaussian according to a Shapiro-Wilk test (adopting the limit $p$-value $<$ 0.001), we calculated the median values and they range from $\sim$0.05 (Mn) to $\sim$0.15 dex (V).

To check if these distributions are consistent with those from the solar neighborhood, we compared them with the ones obtained by \cite{luck17,luck18} from the analysis of a sample of 1777 stars in a ``local region'' within 100 pc of the Sun. Recall that stars in our sample, on the other hand, have distances ranging from $\sim$60 to $\sim$1800 pc. Since they used different solar abundances \citep{scott15a,scott15b,grevesse15}, we put their values on the same reference scale we adopted \citep{asplund09} using the relation [X/H]$_{rescaled}$ = [X/H]$_{L17+L18}$ + A(X)$_{\odot}^{S15+G15}$ - A(X)$_{\odot}^{A09}$, where [X/H]$_{L17+L18}$ are the abundances from \cite{luck17,luck18}, A(X)$_{\odot}^{S15+G15}$ are the absolute solar abundances from \cite{scott15a,scott15b} and \cite{grevesse15}, and A(X)$_{\odot}^{A09}$ are the absolute solar abundances from \cite{asplund09} for a given element X. As their sample spans wider intervals for the atmospheric parameters, we performed the following cuts in their sample: 4750 K $\leq$ \teff $\leq$ 6350 K, 3.40 dex $\leq$ \logg\ $\leq$ 4.90 dex and -0.60 dex $\leq$ \feh\ $\leq$ 0.50 dex. The more restricted sample contains 1368 stars, although not all abundances are available for all stars.

The abundance distributions from \cite{luck17,luck18} are represented by red histograms in Figure \ref{ab_hist}, and we can see a general good overlap between the two samples, but ours does not contain as many stars with lower abundances. Following \cite{ghezzi21}, we investigated whether these samples are different using the Cucconi statistical test\footnote{We used the Python implementation developed by Grzegorz Mika and available at \url{https://github.com/GrzegorzMika/NonParStat}. We chose the option “bootstrap” for the method and 10$^{5}$ replications.}. It simultaneously compares the central tendency and variance of two independent samples and is considered one of the most sensitive nonparametric tests to jointly evaluate differences in location and scale \citep{marozzi13}. As for Fe, the two sets of stars are statistically different ($p <$ 0.001) for all species, except Mg, Si, Sc II, and Cr I. This result reveals that our sample is not representative of the solar neighborhood and this is probably a consequence of the higher metallicities of its stars.

\subsection{Internal consistency checks} 
\label{sec:checks}

When searching for possible connections between different classes of exoplanets and the chemical abundances of their stellar hosts, it is important to guarantee as much as possible that eventual positive results do not arise from underlying correlations in the data. We performed some additional tests to ensure that our abundances were reliable before proceeding with the discussion of the results.

\subsubsection{Possible correlations with stellar parameters} 
\label{sec:correlations}

We investigated the behavior of the abundances as a function of \teff\ and \logg\ through weighted linear fits and found $R^{2} < 0.18$ for \teff\ and $R^{2} < 0.11$ for \logg. Despite these low correlation coefficients, we note that, according to Spearman and Pearson tests, we find significant correlations for the abundances of Na, Mg, Al, Ca, Ti I, V, Cr I, Cr II, Mn, and Cu in the case of \teff\ and for the abundances of Na, Ti I, V, Cr I, Cr II, Mn, Co, and Cu in the case of \logg. 
 
Trends of increasing abundances as a function of decreasing \teff\ have been observed in multiple previous works \citep[e.g.,][]{adibekyan12a,luck17,luck18,brewer18}. An intriguing case is Ti I, which in this study has a good number of lines with moderate to large EWs. We investigated whether this trend could be caused by blended lines, which we evaluated through spectral synthesis and the catalog of \cite{heiter21}, but their removal did not improve the results. We also checked if non-LTE effects could be the cause of the trend, and although they may be significant for Ti lines, the non-LTE corrections for the parameter space of our sample are not high enough to explain the observed trend \citep[e.g.,][]{mallinson22,mallinson24}.

An additional test is to compare the abundances obtained from different ionization stages of the same element. For Ti, we confirm the increase in the Ti I abundances with decreasing \teff, and a similar behavior is observed for Sc I and Cr II. We note that Cr II and Sc I have only seven and four lines, respectively. Together with the typically low S/N values of our spectra, blendings could explain the increasing discrepancies towards lower effective temperatures. For Sc, we also see an increase in the differences for higher effective temperatures, which can be explained by the weakness of the Sc I lines for hotter stars coupled with the typical low S/N values of our spectra. Despite these trends, there is an overall good agreement between the two sets of abundances for each element. The median and MAD values of [X I/H] - [X II/H] are 0.013 $\pm$ 0.079 dex for Sc, -0.020 $\pm$ 0.050 dex for Ti, and -0.045 $\pm$ 0.053 dex for Cr. We note that there are no significant trends of the differences [X I/H] - [X II/H] with \logg. 

\subsubsection{Tests with the solar spectrum} 
\label{sec:sun}

The spectra analyzed in this work have a wide range of S/N values ($\sim$10 -- 300), and here, we describe the tests we performed with the solar-proxy spectrum to check the reliability of our abundances. We first analyzed the spectrum of sunlight reflected off Vesta from \cite{ghezzi18}, which was also observed with the HIRES spectrograph. We measured its S/N value for 143 apparent continuum regions by dividing the average normalized flux in each region by the standard deviation. We then performed seven rounds of 2$\sigma$ clippings until convergence was achieved and found a mean value of S/N = 307, which is close to the value of 315 at 5500 \AA. Using the same methodology applied in \cite{martinez19} and \cite{ghezzi21}, we obtained the following atmospheric parameters: \teff\ = 5792 $\pm$ 16 K, \logg\ = 4.47 $\pm$ 0.06, \feh\ = 0.02 $\pm$ 0.01 dex, and \vmic = 1.014 $\pm$ 0.030 km s$^{-1}$. These are in excellent agreement with the canonical solar values. 

\begin{figure*}
\epsscale{1.0}
\plottwo{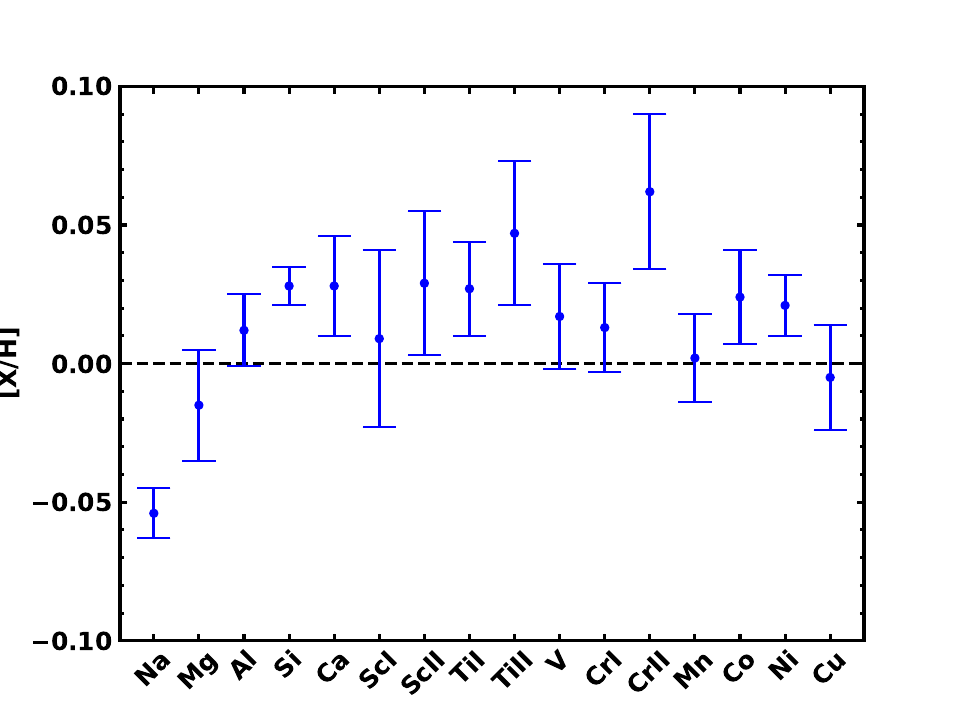}{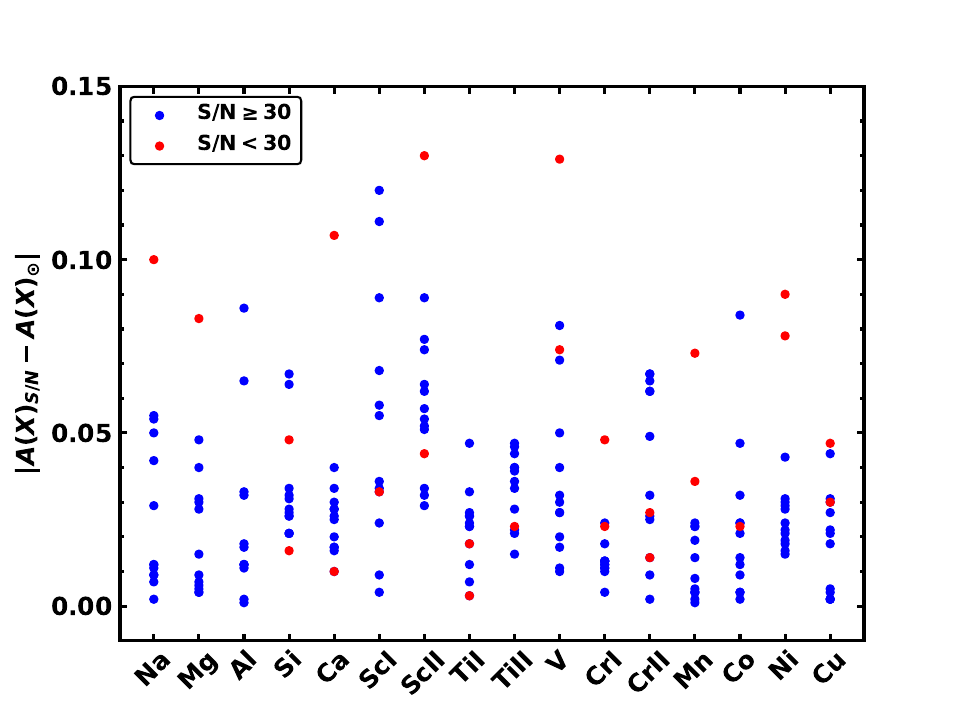}
\caption{\textit{Left panel:} The abundances of 16 species (13 elements, including three - Sc, Ti and Cr - with two ionization stages) determined for the HIRES spectrum of the sunlight reflected off Vesta from \cite{ghezzi18}. \textit{Right panel:} Absolute differences between the abundances determined for the 16 species in the solar spectrum with varying S/N values (300, 250, 200, 150, 100, 90, 80, 70, 60, 50, 40, 30, 20, and 10) and the values from \cite{asplund09}. Points in blue and red represent spectra with S/N $\geq$ 30 and S/N $<$ 30, respectively. Note there are six points with differences larger than 0.15 dex (five for S/N = 10 and one for S/N = 20) not shown in the figure for better visualization purposes.
\label{ab_sun}}
\end{figure*}

Adopting these parameters and following the methodology described in Section \ref{sec:abundances}, we determined the chemical abundances for 16 species (13 elements, including three - Sc, Ti and Cr - with two ionization stages) in the Sun and these can be found in Table \ref{table_param} and the left panel of Figure \ref{ab_sun}. We note that all abundances agree with the solar values from \cite{asplund09} within 2$\sigma$, except for Na, Si and Cr II. This is a consequence of the small internal uncertainties that result from the analysis of a high-quality spectrum of the Sun using solar log \textit{gf} values. The largest absolute difference relative to the reference values adopted from the work of \cite{asplund09} is 0.062 dex for Cr II, with an average $\langle [X/H] \rangle$ = 0.015 $\pm$ 0.026 dex, i.e., in excellent agreement with the solar abundances. 

In order to test the performance of our methodology for lower-quality spectra, we added noise to the solar-proxy spectrum to simulate the following S/N values: 250, 200, 150, 100, 90, 80, 70, 60, 50, 40, 30, 20, and 10. Utilizing the same atmospheric parameters and methodology as above, we obtained new abundances and calculated the absolute differences relative to those from \cite{asplund09}. The results are shown in the right panel of Figure \ref{ab_sun}. We can see that we are able to recover almost all the original abundances within 0.10 dex down to S/N = 30. The only exception is Sc I, which has only a few weak lines in the Sun. Note also that the lower S/N spectra do not necessarily have the larger differences because distinct lines are removed by the clipping procedures, affecting the final average abundances. Below S/N = 30, the average difference for all elements also gets larger than 0.05 dex.

\subsection{Comparison with the literature} 
\label{sec:comp_lit}

\begin{figure*}
\epsscale{1.0}
\plotone{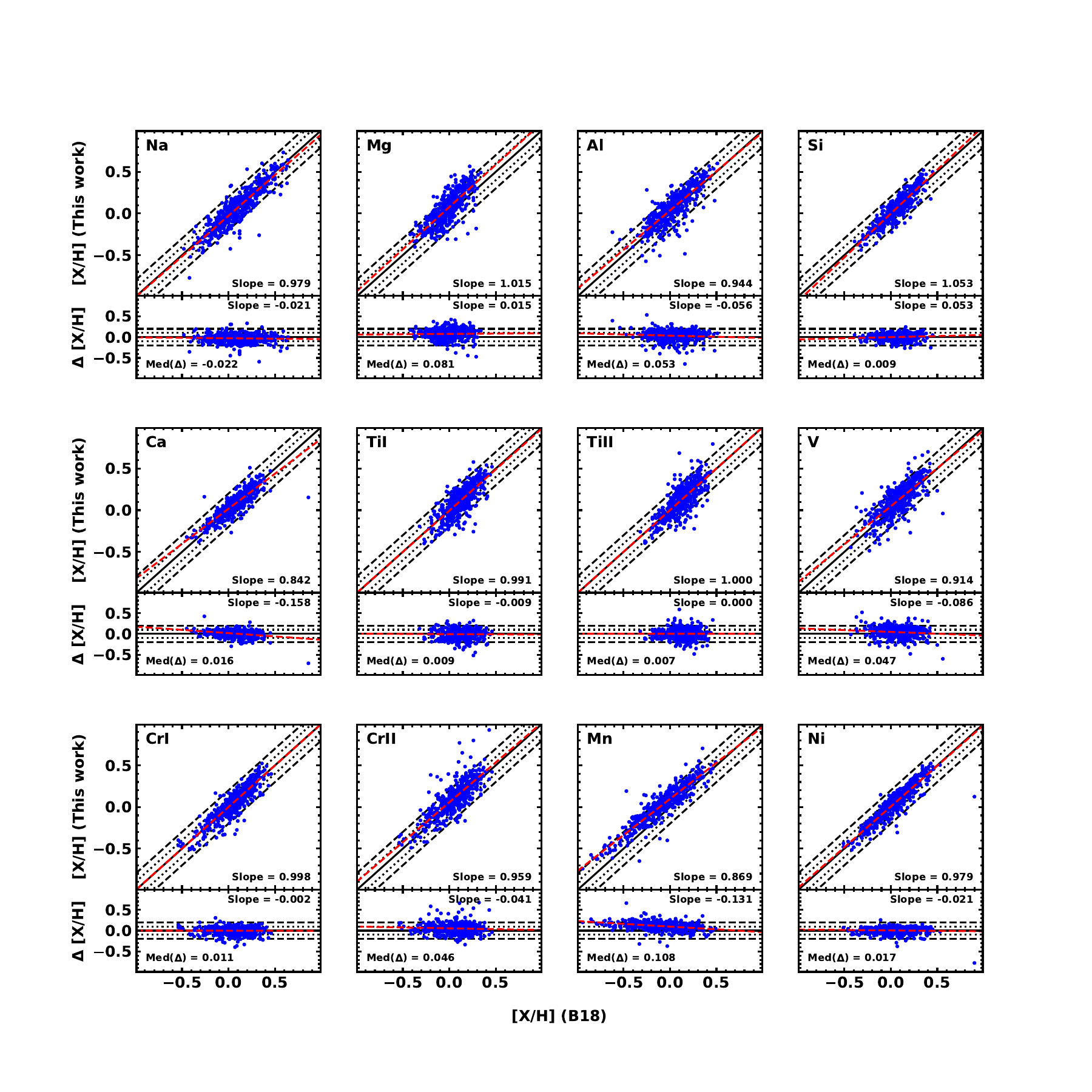}
\caption{Comparison between the abundances from this work and rescaled abundances (see text) from \cite{brewer18} (B18). For each species, the upper panels show the direct comparison between the values, and the lower panels present the differences $\Delta$[X/H] (This work - B18). Note that \cite{brewer18} provide a single set of abundances for Ti and Cr, and we use them in the comparisons with both ionization stages we analyzed for these elements. The solid black lines show a perfect agreement, while dotted and dashed black lines represent, respectively, differences of 0.1 and 0.2 dex as a reference. The red-dashed lines show the linear fits, for which the slopes are shown in each panel, along with the median differences.
\label{comp_b18}}
\end{figure*}

The chemical abundances of the studied elements, except for Sc, Co, and Cu, were previously determined for the CKS stars by \cite{brewer18} using the Spectroscopy Made Easy Software \citep[SME; ][]{pv17} which is based on the spectral synthesis method. They analyzed the same spectra as we did but with a different methodology. Their results are relative to the solar abundances of \cite{gs98}, and thus we applied offsets to put their values on the same reference scale we adopted \citep{asplund09}. In Figure \ref{comp_b18}, we compare our abundances with theirs for 489 stars in common. The distributions of residuals are not Gaussian ($p$-value $<$ 0.001) according to a Shapiro-Wilk test, and the absolute median differences are $\leq$0.05 dex, except for Mg and Mn. The latter has the maximum offset of 0.108 dex. We performed linear fits to residuals and found $R^{2} <$ 0.12 and slopes between -0.158 and 0.053. The Pearson and Spearman tests show that the trends in the residuals are statistically significant for Ca and Mn.  

\section{Discussion} 
\label{sec:discussion}
 
In the following discussions, we will compare the abundance distributions from different subsamples, so we removed evolved stars with \logg\ $<$ 4.0 to minimize any possible evolutionary effects. Moreover, considering the tests in Section \ref{sec:sun}, we decided to cut stars that had spectra with S/N $<$ 30 to guarantee we did not include less reliable abundances. Finally, we removed 10 stars for which it was not possible to determine ages. After these cuts, our sample had 510 stars.   

\subsection{Abundance patterns for the stellar sample} 
\label{sec:stars}

\begin{figure*}
\epsscale{1.0}
\plotone{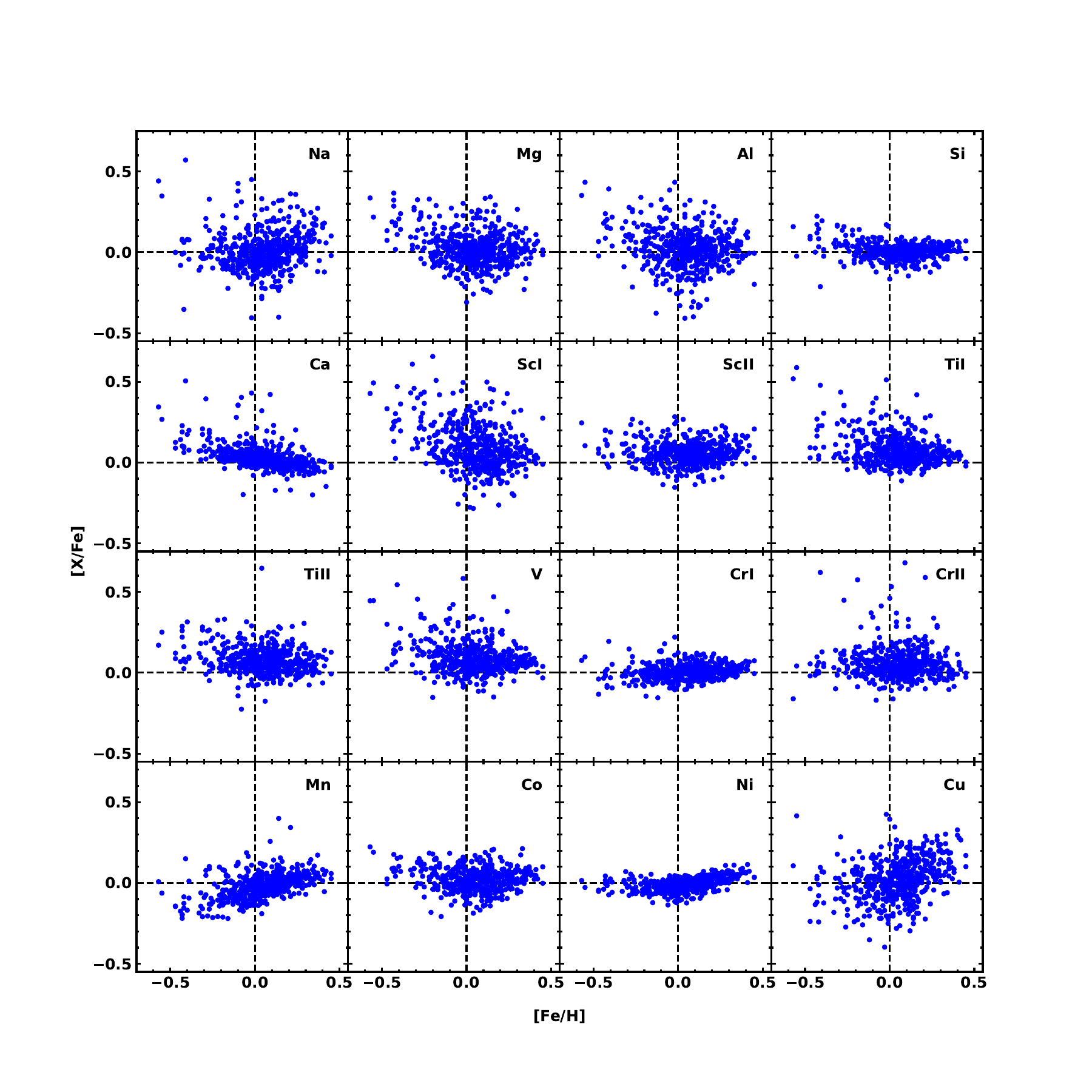}
\caption{Chemical abundances [X/Fe] of 16 species (13 elements, including three - Sc, Ti and Cr - with two ionization stages) as a function of [Fe/H] for the 510 stars in our work. The dashed black lines show the solar values for reference.
\label{ab_xfe_feh}}
\end{figure*}

The abundances of our stars are shown with the canonical diagrams [X/Fe] versus \feh\ in Figure \ref{ab_xfe_feh}. We can see that almost all elements follow their expected behaviors, with iron-peak elements (such as Cr, Co, and Ni) having an almost constant abundance around [X/Fe] = 0 and the $\alpha$-elements (Mg, Si, Ca, and Ti) exhibiting an increase toward lower metallicities. The only exception is Mn, which shows a slight increase with [Fe/H], also seen by \cite{wilson22}.

\begin{figure}
\epsscale{1.0}
\plotone{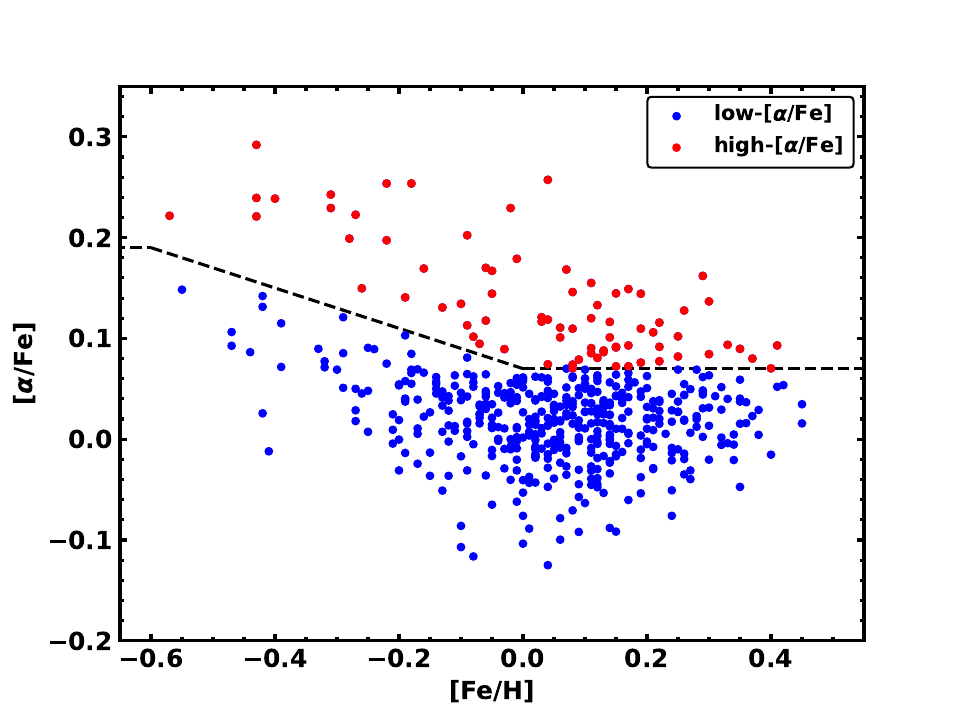}
\caption{Average abundances of $\alpha$-elements [$\alpha$/Fe] (calculated considering the abundances of Mg, Si, and Ti, as in \citealt{adibekyan11}) for our sample. The dashed black line shows to the approximate separation (see text) between low- (blue) and high-[$\alpha$/Fe] (red) stars proposed by \cite{adibekyan11}.
\label{ab_alphafe_feh}}
\end{figure}

One interesting feature in Figure \ref{ab_xfe_feh} is the possible presence of $\alpha$-enhanced stars \citep{adibekyan11,adibekyan12a} in our sample. To quantitatively assess this, in Figure \ref{ab_alphafe_feh}, we show the [$\alpha$/Fe] abundance ratio (average of Mg, Si, and Ti, as in \citealt{adibekyan11}) versus [Fe/H] for our sample along with the separation between the low- and high-$\alpha$ sequences estimated from Figure 1 of \cite{adibekyan11}. For \feh\ $<$ -0.6, [$\alpha$/Fe] = 0.19; for \feh\ $>$ 0.0, [$\alpha$/Fe] = 0.07; and for -0.6 $\leq$ \feh\ $\leq$ 0.0, [$\alpha$/Fe] = -0.20 $\times$ \feh\ + 0.07.

\begin{figure}
\epsscale{1.0}
\plotone{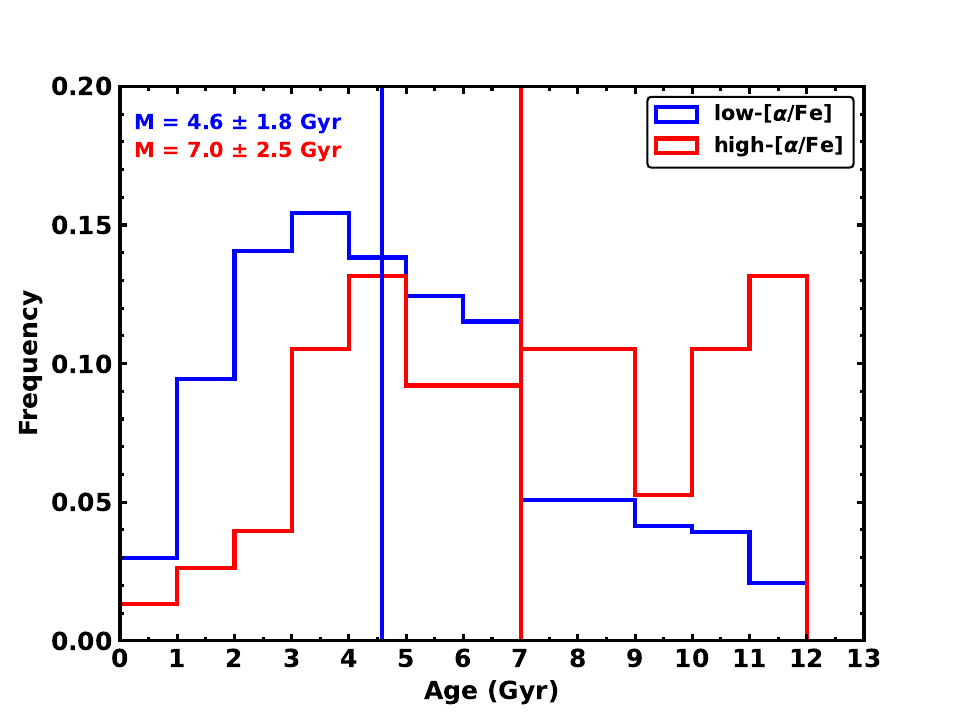}
\caption{Distributions of ages for the low- (blue) and high-[$\alpha$/Fe] (red) stars. The solid vertical blue and red lines represent the respective median values (M), which are also shown along with the corresponding MAD values.
\label{hist_ages}}
\end{figure}

According to this separation, our sample of Kepler planet hosts has 76 stars on the high-[$\alpha$/Fe] sequence and 434 stars on the low-[$\alpha$/Fe] sequence. The age distributions for these two samples are shown in Figure \ref{hist_ages}. The ages of the high-[$\alpha$/Fe] stars range from $\sim$0.92 to 11.50 Gyr, with a median value of 7.0 $\pm$ 2.5 Gyr. Low-[$\alpha$/Fe] stars have ages between $\sim$0.24 and 11.50 Gyr, with a lower median value of 4.6 $\pm$ 1.8 Gyr, which is similar to the age of the Sun ($\sim$4.5 Gyr). The Cucconi test shows that these two samples of stars on the low- and high-$\alpha$ sequences have statistically different distributions of ages ($p$-value = 10$^{-5}$). The median distance of the high-$\alpha$ stars is 742 $\pm$ 143 pc, while the low-[$\alpha$/Fe] stars have a smaller median value of 575 $\pm$ 187 pc. As for the age, the distributions of distances are statistically different ($p$-value = 4 $\times 10^{-5}$). Therefore, high-$\alpha$ stars are typically older and more distant than their low-$\alpha$ counterparts, as generally expected for a population of thick versus thin disks of the Galaxy.

\begin{figure*}
\epsscale{1.0}
\plotone{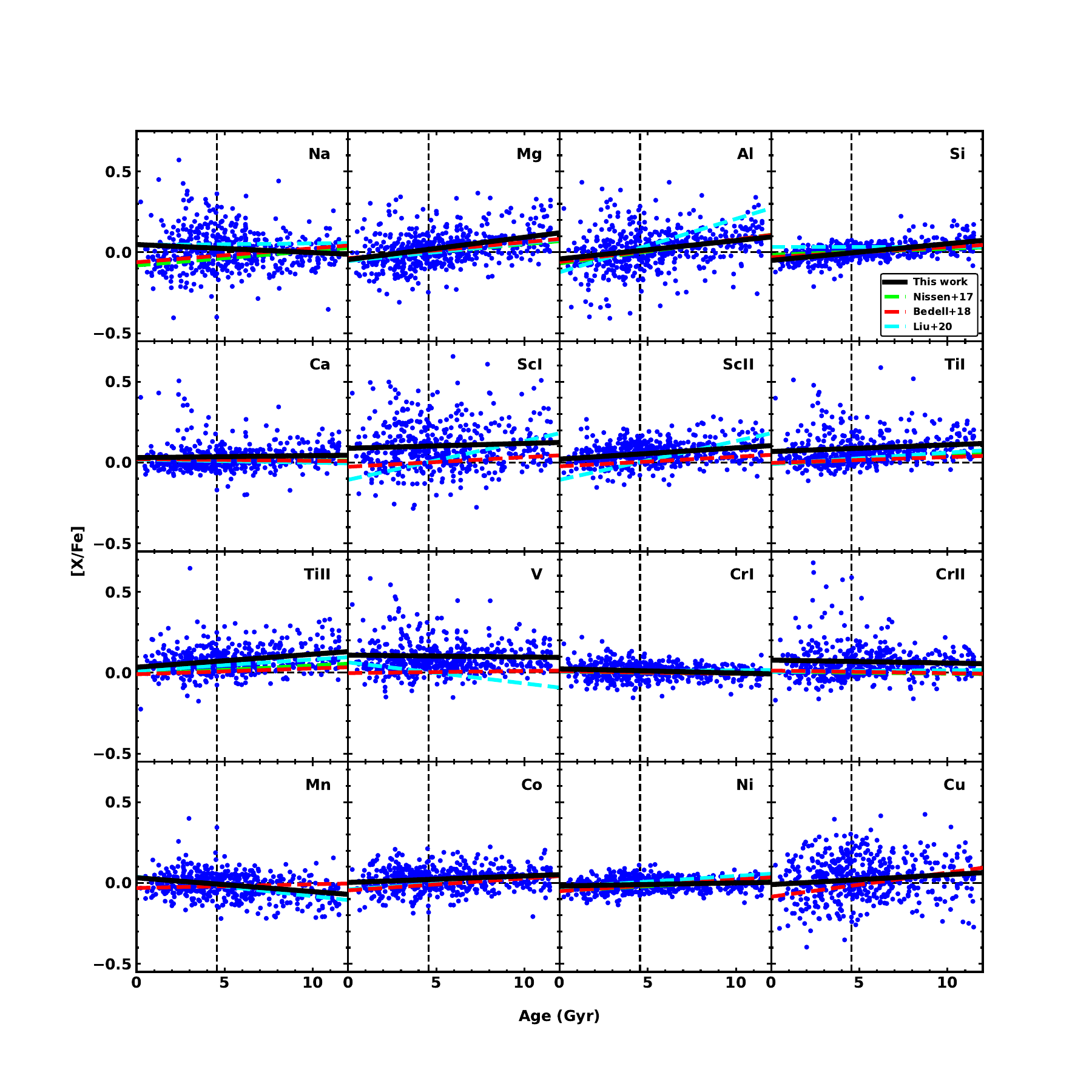}
\caption{Chemical abundances [X/Fe] of 16 species (13 elements, including three - Sc, Ti and Cr - with two ionization stages) as a function of ages for the 510 stars in our work. The dashed black lines show the solar values for reference. The solid black lines represent the weighted least-squares (WLS) fits determined in this work, with the coefficients given in Table \ref{table_gce}. The green-, red- and cyan-dashed lines show the relations between abundances and ages determined by \cite{nissen17}, \cite{bedell18} and \cite{liu20}, respectively.
\label{ab_xfe_age}}
\end{figure*}

Since the abundances vary over time due to Galactic Chemical Evolution (GCE), as can be seen in Figure \ref{ab_xfe_age}, we also investigated whether there are any correlations between their values and stellar ages. We performed weighted least-squares (WLS) fits (using the uncertainties on the abundances as the weights) and determined the coefficients $a$ and $b$ for the equation [X/Fe] = (a$_{X}$ $\times$ Age) + b$_{X}$ for each element X (see Table \ref{table_gce}), where [X/Fe] = [X/H] - [Fe/H]. According to Pearson and Spearman tests, we have statistically significant correlations for Mg, Al, Si, Ca, Sc II, Ti I, Ti II, Mn, and Co. The stronger relation is observed for Mg, which shows a slope of 0.014 dex Gyr$^{-1}$. In Figure \ref{ab_xfe_age}, we can see that the relations determined in this work are in good agreement with those obtained by \cite{nissen17}, \cite{bedell18}, and \cite{liu20}, despite the differences between the samples and methodologies adopted by the different studies. 

\begin{deluxetable}{lrrrr}
\tablecaption{Coefficients for the weighted linear fits between [X/Fe] and age. 
\label{table_gce}
}
\tablehead{\colhead{Species} & \colhead{\textit{a}} & \colhead{\textit{b}} & \colhead{$p$-value (Spearman)} & \colhead{$p$-value (Pearson)} \\
\colhead{} & \colhead{(dex Gyr$^{-1}$)} & \colhead{(dex)} & \colhead{} & \colhead{}
}
\startdata
Na & -0.005$\pm$0.002 & 0.048$\pm$0.012 & $9.442 \times 10^{-1}$ & $3.147 \times 10^{-1}$ \\
Mg & 0.014$\pm$0.002 & -0.045$\pm$0.010 & $2.463 \times 10^{-19}$ & $1.353 \times 10^{-18}$ \\
Al & 0.011$\pm$0.002 & -0.041$\pm$0.012 & $1.045 \times 10^{-15}$ & $8.339 \times 10^{-12}$ \\
Si & 0.010$\pm$0.001 & -0.050$\pm$0.004 & $6.005 \times 10^{-32}$ & $2.375 \times 10^{-36}$ \\
Ca & 0.001$\pm$0.001 & 0.030$\pm$0.009 & $1.155 \times 10^{-8}$ & $5.419 \times 10^{-3}$ \\
Sc I & 0.003$\pm$0.002 & 0.088$\pm$0.014 & $3.074 \times 10^{-1}$ & $2.598 \times 10^{-1}$ \\
Sc II & 0.007$\pm$0.001 & 0.021$\pm$0.006 & $5.469 \times 10^{-10}$ & $5.350 \times 10^{-10}$ \\
Ti I & 0.004$\pm$0.002 & 0.068$\pm$0.010 & $3.243 \times 10^{-11}$ & $3.886 \times 10^{-5}$ \\
Ti II & 0.008$\pm$0.001 & 0.034$\pm$0.008 & $7.728 \times 10^{-9}$ & $2.077 \times 10^{-9}$ \\
V & -0.001$\pm$0.002 & 0.109$\pm$0.010 & $1.642 \times 10^{-1}$ & $8.866 \times 10^{-1}$ \\
Cr I & -0.003$\pm$0.001 & 0.024$\pm$0.005 & $9.937 \times 10^{-2}$ & $1.047 \times 10^{-2}$ \\
Cr II & -0.002$\pm$0.002 & 0.078$\pm$0.011 & $1.121 \times 10^{-2}$ & $9.813 \times 10^{-1}$ \\
Mn & -0.009$\pm$0.001 & 0.032$\pm$0.007 & $2.944 \times 10^{-9}$ & $4.182 \times 10^{-12}$ \\
Co & 0.004$\pm$0.001 & 0.003$\pm$0.007 & $2.300 \times 10^{-3}$ & $2.278 \times 10^{-3}$ \\
Ni & 0.002$\pm$0.001 & -0.019$\pm$0.004 & $5.030 \times 10^{-3}$ & $1.506 \times 10^{-2}$ \\
Cu & 0.006$\pm$0.002 & -0.010$\pm$0.012 & $1.660 \times 10^{-2}$ & $2.072 \times 10^{-2}$ \\
\enddata
\end{deluxetable}

\subsection{Stellar abundances as a function of exoplanet properties} 
\label{sec:planets}

\begin{figure*}
\epsscale{1.0}
\plotone{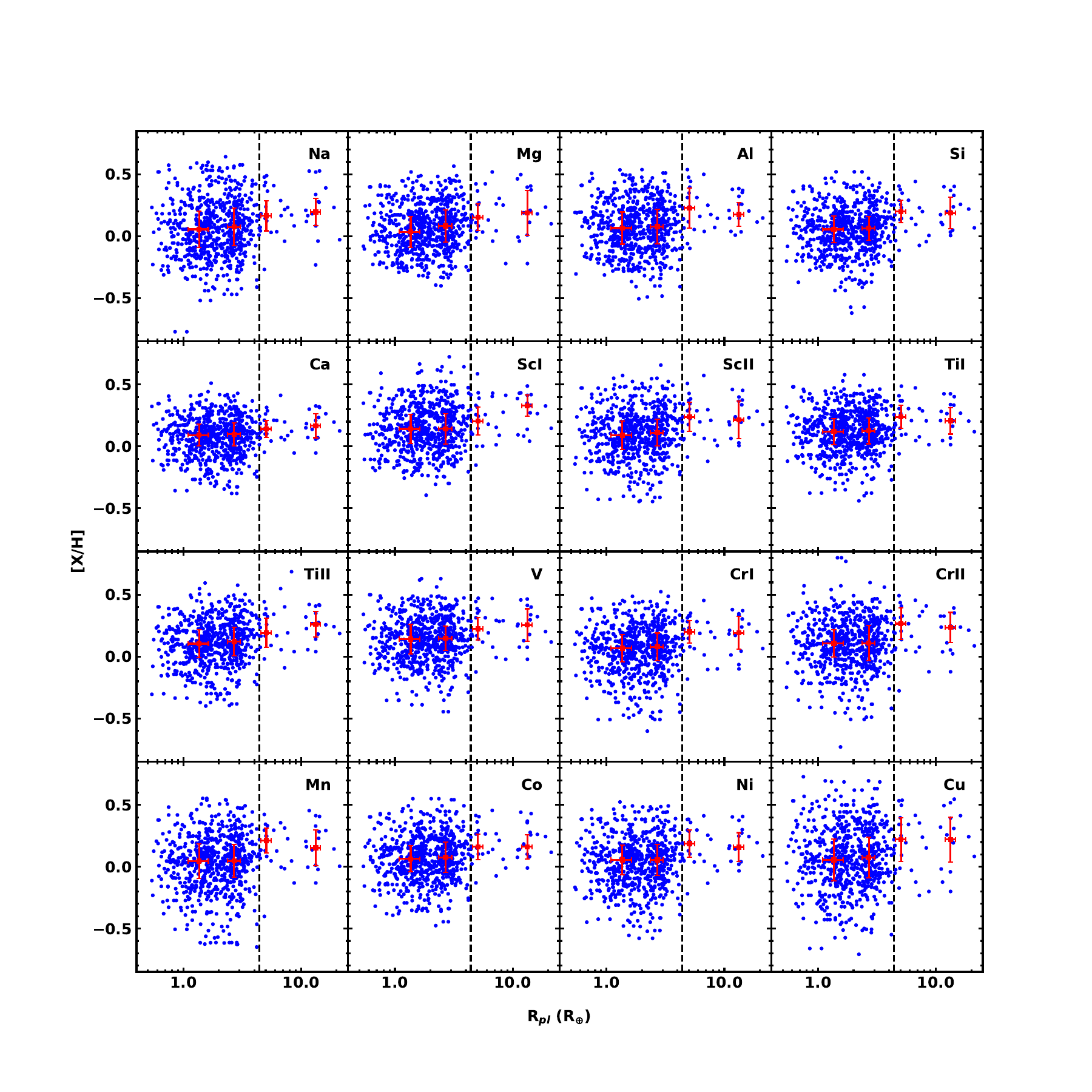}
\caption{Abundances [X/H] of all elements as a function of planetary radii. The red squares represent the median values for each class of planets (SE, SN, SS, and JP, in order of increasing radius), and the error bars are the median absolute deviations.
\label{ab_xh_rpl}}
\end{figure*}

The 510 stars in our sample host 726 planets: 367 super-Earths (SE), 318 sub-Neptunes (SN), 23 sub-Saturns (SS) and 18 Jupiters (JP). In Figure \ref{ab_xh_rpl}, we show the derived stellar abundances [X/H] (shown as filled blue circles) as a function of the exoplanetary radii, with the latter taken from \cite{martinez19}. We note that in this figure the abundances from hosts of multiplanetary systems are plotted individually and thus appear more than one time. We find that the median values (shown as filled red circles) are higher for stars with large planets (SS or JP) relative to those hosting small planets (SE or SN) for all elements, and this result is consistent with the one found by \cite{ghezzi21} for Fe. The minimum and maximum differences between the median abundances are 0.055 dex for Ca and 0.155 dex for Cu, respectively. The median difference for all elements is 0.123 $\pm$ 0.008 dex. The Pearson test shows that there are significant correlations ($p$-value $<$ 0.001) for Na, Mg, Al, Si, Sc I, Sc II, Ti I, Ti II, Cr I, Mn, Co, and Ni. The Spearman test, on the other hand, returns $p$-values $<$ 0.001 for Mg, Sc II, Ti II, Mn, and Co. We also performed Cucconi tests and found that the distributions of abundances for stars with large or small planets are significantly different ($p$-value $<$ 0.001) for all 16 species, except Ca, V, and Cu. 

If we use [X/Fe] abundances instead, the median values are similar for the four classes of planets (SE, SN, SS, and JP), suggesting that there are no significant correlations with planet size. This is confirmed by Spearman and Pearson tests. These results agree with those from \cite{wilson22} for the elements analyzed in both works (Mg, Al, Si, Ca, Mn, and Ni), except for [Mn/Fe], for which they find a correlation, and we do not. However, \cite{wilson22} consider that their result for Mn was probably caused by a strong underlying correlation between [Mn/Fe] and [Fe/H] in the APOGEE abundance results (which we also observe in our sample). The absence of correlations between [X/Fe] and planet radii suggests that the trends observed for [X/H] reflect the underlying correlation between [Fe/H] and R$_{pl}$.

We also investigated trends with orbital periods (taken from \citealt{martinez19}) by dividing our sample in stars with hot (P $\leq$ 10 days; N = 395) and warm (10 days $<$ P $\leq$ 100 days; N = 331) exoplanets. The median difference between stars with hot and warm planets is 0.050 $\pm$ 0.009 for all elements, with a minimum value of 0.019 for Sc I and a maximum value of 0.076 for Al. We find statistically significant correlations ($p$-value $<$ 0.001) for Ca and Cu in the Pearson test and for all species except Mg, Sc I, and Ti II in the Spearman test. Moreover, the Cucconi tests reveal that the distributions of abundances for stars with hot or warm planets are significantly different ($p$-value $<$ 0.001) for all 16 species, except Mg, Sc I, Sc II, Ti II, and Ni. As before, if we use [X/Fe] abundances instead, we do not find $p$-value $<$ 0.001 for any of the species. This result also agrees with the conclusions of \cite{wilson22} for all elements in common (Mg, Al, Si, Ca, Mn, and Ni) and supports the hypothesis the trends observed for [X/H] simply reflect a dependence between orbital period and [Fe/H].

\subsection{Abundance patterns for exoplanetary systems} 
\label{sec:systems}

In Section \ref{sec:abundances}, we discussed that the abundances for Sc I rely on a few lines and their errors are typically higher than for Sc II. For Cr, we have the opposite situation, and the uncertainties for Cr II are larger on average. In Section \ref{sec:correlations}, we showed that the abundances of Ti I present a significant correlation with \teff, which is also seen in other works in the literature. In Figure \ref{ab_xfe_feh}, we can see that the dispersions for Sc I, Ti I, and Cr II are larger relative to Sc II, Ti II, and Cr I. Finally, the discussion in Section \ref{sec:planets} showed that results are not always consistent for the two ionization stages. Thus, in the following discussions, we will adopt the more reliable abundances obtained from Sc II, Ti II, and Cr I for Sc, Ti, and Cr, respectively.

The analysis of individual planets might be affected by the multiplanetary systems since the abundances of their host stars are considered more than once in the statistical tests. To avoid this, here, we divided our sample of 510 stars according to the classifications of planetary systems described in Section \ref{sec:data}. For the following tests, we compared the abundance distributions of the subsamples using the Cucconi test and considering that differences were statistically significant when $p$-value $<$ 0.001. Table \ref{table_stat_tests} summarizes the results of these comparisons and shows only the elements for which we found statistically significant differences in each comparison.

\begin{deluxetable}{lcc}
\tablecaption{Elements for which statistically significant differences were found in the Cucconi tests ($p$-value $<$ 0.001), with Sc, Ti, and Cr represented by the species Sc II, Ti II, and Cr I, respectively.
\label{table_stat_tests}
}
\tablehead{\colhead{Samples} & \colhead{[X/H]} & \colhead{[X/Fe]}
}
\startdata
Small (472) -- Large (27) & Al, Si, Sc, Ti, Cr, Fe, Co & -- \\
Hot (241) -- Warm (192) & Al, Si, Ca, Fe, Co, Ni & -- \\ 
Small H (223) -- Small W (179) & Si, Fe, Co & -- \\
Large H (16) -- Large W (10) & -- & -- \\
SE (286) -- SN (192) & -- & -- \\
SE H (163) -- SE W (40) & -- & -- \\
SN H (42) -- SN W (130) & -- & -- \\
SE H (163) -- SN H (42) & -- & -- \\
SE W (40) -- SN W (130) & -- & -- \\
SE+SN H (18) -- SE+SN W (9) & -- & -- \\
Small single (330) -- Small multi (142) & -- & -- \\
SE single (176) -- SE multi (45) & -- & -- \\
SN single (154) -- SN multi (32)  & -- & -- \\
SE single H (140) -- SE single W (36)  & -- & -- \\
SE multi H (23) -- SE multi W (4) & -- & -- \\
SN single H (38) -- SN single W (116)  & -- & -- \\
SN multi H (4) -- SN multi W (14) & -- & -- \\
\enddata
\end{deluxetable}

\subsubsection{Planet Size, Orbital Period and Multiplicity}
\label{sec:radii}

\begin{figure}
\epsscale{1.0}
\plotone{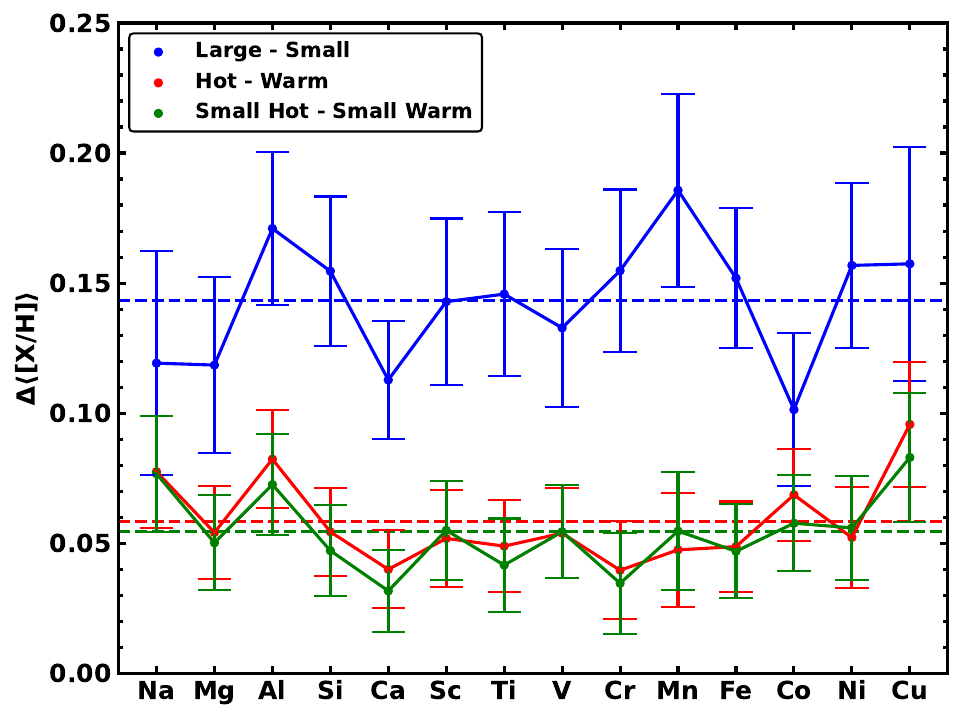}
\caption{Differences between the average abundances for systems having only large and only small planets (blue), systems having only hot and only warm planets (red), and systems having only small hot and only small warm planets (green). The error bars show the standard deviations of the mean for the abundance differences. The dashed lines represent the average differences for each of the three comparisons.
\label{diff_avg_systems}}
\end{figure}

The first comparison we performed was between systems with only small (N = 472) and only large (N = 27) planets, and we found that, as for Fe, the abundances [X/H] for all elements are systematically higher for the latter. In particular, our result for [Mg/H] agrees with the one found by \cite{lt25}. The differences between the average abundances are shown in blue in Figure \ref{diff_avg_systems}, and the mean and median values for all elements are 0.143 $\pm$ 0.024 dex and 0.135 $\pm$ 0.011 dex, respectively. This result is a direct consequence of the well-known planet-metallicity correlation, which is valid only for the overall population of large planets \citep[e.g.,][]{fischer05,buchhave12,petigura18,ghezzi21,wilson22}. The Cucconi test shows that the [Fe/H] distributions are statistically different and the same holds for the elements Al, Si, Sc, Ti, Cr, and Co (see Table \ref{table_stat_tests}).

We then compared systems with only hot (N = 241) and only warm (N = 192) planets and observed a statistically significant difference for Al, Si, Ca, Fe, Co, and Ni (see Table \ref{table_stat_tests}). These results are consistent with the higher abundances found for stars with hot planets for iron \citep[e.g.,][]{mulders16,petigura18,wilson18}, as well as for other elements \citep{wilson22}, except for Mg and Mn. In Figure \ref{diff_avg_systems}, we show the differences between the average abundances in red, and it is clear that systems with hot planets are systematically more enriched in metals, with mean and median differences of 0.058 $\pm$ 0.017 dex and 0.057 $\pm$ 0.005 dex, respectively.  

The next comparison was between systems with only small hot (N = 223) and only small warm (N = 179) planets. We find that the subsamples are statistically different for Si, Fe, and Co (see Table \ref{table_stat_tests}). The mean and median differences are 0.054 $\pm$ 0.015 dex and 0.060 $\pm$ 0.011 dex, respectively, which are very similar to the values found in the previous test (see also Figure \ref{diff_avg_systems}), as expected due to the fact that our sample is mainly composed of systems with only small planets. For the comparison between systems with only large hot (N = 16) and only large warm (N = 10) planets, we found no statistically significant differences. 

The small sample sizes for the large planets prevent further divisions into subsamples. For this reason, we now take a closer look only at the sample of systems with small planets, dividing it into systems that contain only super-Earths (SE; N = 286) and sub-Neptunes (SN; N = 192). The comparison between systems hosting only SE and only SN yields no statistical differences for any of the elements. A similar result was found when comparing SE H (N = 163) and SE W (N = 40), SN H (N = 42) and SN W (N = 130), SE H and SN H, SE W and SN W, and SE+SN H (N = 18) and SE+SN warm (N = 9) (see Table \ref{table_stat_tests}). 

The absence of differences is intriguing since \cite{wilson22} found distinct occurrence rates for hot and warm super-Earths as well as for hot and warm sub-Neptunes for all 10 elements analyzed, including 7 elements in common with our work (Mg, Al, Si, Ca, Mn, Fe, and Ni). However, we observe that SE H systems have typically larger abundances relative to SE W systems (mean and median differences are 0.097 $\pm$ 0.023 dex and 0.099 $\pm$ 0.014 dex, respectively). The SN H systems are also enriched relative to SN W systems, although the mean and median differences are smaller (0.050 $\pm$ 0.017 dex and 0.047 $\pm$ 0.020 dex, respectively). These results are consistent with the findings of \cite{wilson22} that differences in the occurrence rates of hot and warm planets are more pronounced for super-Earths relative to sub-Neptunes. Therefore, the fact that we did not observe any statistically significant differences in the Cucconi tests could result from different samples, classification of the planets (\citealt{wilson22} uses the limit 4.0 R$_{\oplus}$ instead of 4.4 R$_{\oplus}$ to divide SN and SS), consideration of planetary systems as opposed to individual planets, and the stellar abundances. 

We also considered the possible influence of multiplicity by comparing the following pairs of systems: small single (N = 330) versus small multi (N = 142), SE single (N = 176) versus SE multi (N = 45) and SN single (N = 154) versus SN multi (N = 32). We did not find any statistical differences for any of the elements (see Table \ref{table_stat_tests}), and these results are consistent with previous findings for Fe from \cite{weiss18} and \cite{ghezzi21} and for Mg from \cite{lt25}. Finally, we considered simultaneously the radii and orbital periods of the planets as well as the multiplicity in the system with the following comparisons: SE single H (N = 140) versus SE single W (N = 36), SE multi H (N = 23) versus SE multi W (N = 4), SN single H (N = 38) versus SN single W (N = 116) and SN multi H (N = 4) versus SN multi W (N = 14). As before, we did not find statistically significant differences for any of the elements (see Table \ref{table_stat_tests}). It is interesting to note that the difference observed by \cite{ghezzi21} between SE single H and SE single W for Fe has not been recovered here ($p$-value = 0.036). Although we use the same metallicities as in \cite{ghezzi21} for the studied stars, we recall that the sample discussed in that study was larger and included metallicities obtained by \cite{petigura17} for those stars having low signal-to-noise HIRES spectra or large projected rotational velocities. 
 
We also checked whether considering [X/Fe] (corrected or not for the trends with age) instead of [X/H] would change the results. Now, we do not find any statistical differences in the comparisons between systems with small and large planets (consistent with the results for Mg from \citealt{lt25}) as well as small hot and small warm planets (see Table \ref{table_stat_tests}), supporting the conclusion that the differences observed for [X/H] simply reflect the planet-metallicity correlation. For the comparison between systems with only hot and only warm planets, we do not see statistical differences for all elements either, and this result supports the conclusion that the differences found when using [X/H] are caused by the underlying influence of [Fe/H] in the occurrence rates of hot exoplanets. For the remaining comparisons, we see no statistical differences. In particular, the absence of a statistically significant difference in the comparison of [Mg/Fe] distributions for small single and small multi systems is consistent with the result found by \cite{lt25}. 

Since the individual abundances did not reveal any clear patterns with the architectures of planetary systems, we also investigated them together using four different quantities: median [X/H], median [X/Fe], median [X/Fe] corrected for age trends and [Ref/H] = A(Ref)$_{\star}$ - A(Ref)$_{\odot}$, where Ref means refractory elements, A(Ref) = $\log (\sum_{X}$ 10$^{A(X)}$), A(X)$_{\star}$ = [X/H]$_{\star}$ + A(X)$_{\odot}$ and A(X)$_{\odot}$ are the solar abundances for a given element X from \cite{asplund09}. The results are very similar to those presented in Table \ref{table_stat_tests}, and we only found statistically significant differences for the median [X/H] in the comparisons Small - Large, Hot - Warm and Small H - Small W as well as for [Ref/H] in the first of these cases. 

Specific groups of elements have different roles in the planetary formation processes (e.g., \citealt{hinkel19} and \citealt{torres-quijano25}). Namely, the lithophiles are found in rock-forming minerals important to planetary interiors, while the siderophiles combine with iron and are present in the planetary cores. We investigated if the abundances of these two groups correlate with the planetary architectures. Following \cite{hinkel19} and \cite{torres-quijano25}, we considered the groups of lithophiles (Na, Mg, Al, Si, Ca, Sc, Ti, V, and Mn) and siderophiles (Cr, Co, and Ni), and, as for the refractories, we tested both their median and summed abundances. In all four cases, we found a significant difference between systems with only small and only large planets. For the comparisons Hot - Warm and Small H - Small W, there is a statistical difference for the median abundance of the lithophiles, which is not corroborated by their summed abundances.

For completeness, we also performed the comparisons listed in Table \ref{table_stat_tests} for [$\alpha$/Fe] and the abundance ratios Fe/Si and Mg/Si. We did not find any statistically significant differences; however, the abundances of the $\alpha$-elements presented some interesting features that deserve a more detailed discussion.

\subsubsection{Low- and high-$\alpha$ systems}
\label{sec:alpha}
 
\begin{figure}
\epsscale{1.0}
\plotone{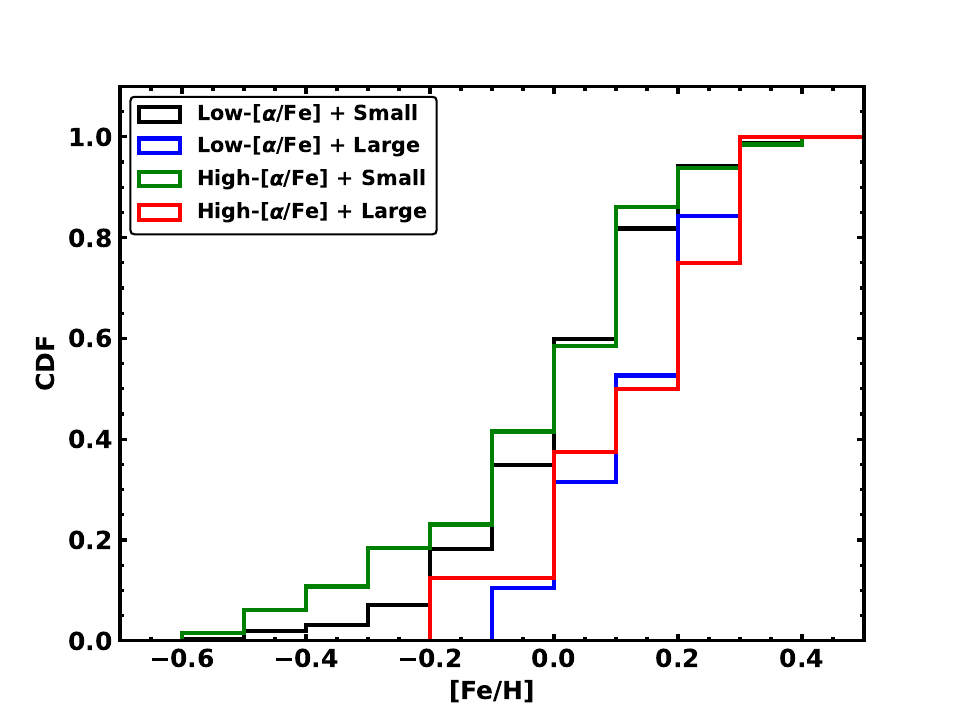}
\caption{Cumulative distributions functions (CDFs) for low-[$\alpha$/Fe] stars hosting only small (black) or large (blue) planets and for high-[$\alpha$/Fe] stars hosting only small (green) or large (red) planets. 
\label{cdfs_low_high_alpha}}
\end{figure}

In Section \ref{sec:stars}, we showed that our sample has 76 and 434 high- and low-[$\alpha$/Fe] stars, respectively. The latter sample has 407 systems with only small planets and 19 with only large planets. Although the maximum metallicities for these two groups are similar (0.45 and 0.36 dex, respectively), the minimum value is higher for systems with only large planets (-0.03 dex) relative to systems with only small planets (-0.55 dex). These results are displayed in Figure \ref{cdfs_low_high_alpha} and reflect the well-known giant planet-metallicity correlation. 

Among the high-[$\alpha$/Fe] stars, 65 have only small planets and 8 have only large planets. As for the low-[$\alpha$/Fe] stars, the maximum metallicities for the two samples are similar (0.40 and 0.35 dex, respectively), but the minimum value is larger for high-[$\alpha$/Fe] stars with only large planets (-0.19 dex) relative their counterparts with only small planets (-0.57 dex). These results can also be seen in Figure \ref{cdfs_low_high_alpha}. 

Interestingly, the minimum \feh\ value is similar for low- and high-$\alpha$ systems with only small planets. However, it is higher (by 0.16 dex) for low-$\alpha$ systems with only large planets relative to their high-$\alpha$ counterparts. In order to test the significance of this last result, we performed a Monte Carlo resampling of each stellar metallicity within its respective uncertainty considering both Gaussian and uniform distributions. For 10,000 realizations, the median differences are 0.15 $\pm$ 0.02 dex and 0.16 $\pm$ 0.01 dex, respectively. Therefore, the difference is statistically significant, and it seems that higher abundances of the $\alpha$-elements are able to partially compensate for the lower iron abundances, allowing the formation of giant planets in somewhat more metal-poor environments. Although these results are consistent with the findings of \cite{adibekyan12a,adibekyan12b}, the higher minimum metallicity for the high-[$\alpha$/Fe] systems with only large planets relative to their counterparts with only small planets suggests that, even if the abundance of $\alpha$-elements is higher, the Fe abundance still acts as a limiting factor in the formation of large exoplanets. We note, however, that the Cucconi test shows no significant differences between the metallicity distributions of low- and high-$\alpha$ systems with only large planets.

\cite{adibekyan12b} explored the metallicity interval -0.65 $<$ \feh\ $<$ -0.30 and found that stars hosting exclusively small planets were typically more enhanced in $\alpha$-elements. Our sample has 20 systems in this low-metallicity regime, and all of them have only small planets. Since 13 of them (65\%) belong to the low-$\alpha$ sequence, our results support the conclusion that an enhancement in the overall metallicity is not a requirement for the formation of small planets \citep{buchhave12}. Note, however, that \cite{adibekyan12b} used just the Ti abundance in his analysis while we adopted the average of Mg, Si, and Ti (see Section \ref{sec:stars}). If we consider only the Ti abundance, 11 systems (55\%) belong to the low-$\alpha$ sequence, which confirms our previous conclusion.  

We also compared the metallicity distributions for low-[$\alpha$/Fe] stars hosting only hot and only warm planets and  did not find a statistically significant difference. A similar result was obtained for the high-[$\alpha$/Fe] systems. Moreover, we calculated the median maximum radius for planets orbiting low-[$\alpha$/Fe] stars (2.07 $\pm$ 0.75 R$_{\oplus}$) and noted that it is slightly smaller than for planets around their high-[$\alpha$/Fe] counterparts (2.59 $\pm$ 0.91 R$_{\oplus}$). However, the difference in the distributions of the maximum radius is not statistically significant. Finally, we also performed Cucconi tests to compare the maximum orbital periods and the number of planets in the systems, but found no statistically significant differences. 
 
\subsection{Chemical abundance versus condensation temperature trends} 
\label{sec:slopes}

In the previous section, we analyzed the individual abundances of each element and concluded that, in general, they closely follow the well-known planet-metallicity correlation. However, it has previously been suggested in the literature that abundance patterns for groups of elements could be related to the formation of different classes of planets \citep[e.g.,][]{hinkel19,torres-quijano25}. In particular, the slopes of the abundances of refractory elements (T$_{C} >$ 900 K) as a function of condensation temperatures could result from the formation of terrestrial or giant planets \citep[e.g.,][]{melendez09,bedell18,booth20}. However, more recent studies discuss that this chemical signature that was originally found for the Sun does not seem to be related to planetary formation (\citealt{liu20,rampalli24,carlos25,rampalli25};  see also the review by \citealt{gustafsson25}).  

\subsubsection{Complete sample} 
\label{sec:slopes_all}

Our sample has 510 stars hosting 726 planets, among which we have 472 systems with only small planets and 286 systems with only super-Earths. Therefore, this large and diverse sample offers a unique opportunity to test whether the formation of terrestrial planets imprints chemical signatures on solar-type stars. Using the derived abundances [X/Fe], the respective uncertainties determined in this work and the 50\% condensation temperatures $T_{C}$ for the elements from \cite{lodders25}, we performed weighted least squares fits (WLS) to the data and determined the slopes and associated statistics for each host star; these are presented in Table \ref{table_param}. Here, we use only [X/Fe] abundances (instead of [X/H]) to minimize the effects of atomic diffusion due to the different ages and masses of stars \citep[e.g.,][]{dotter17,bedell18}. 

\begin{figure*}
\epsscale{1.0}
\plottwo{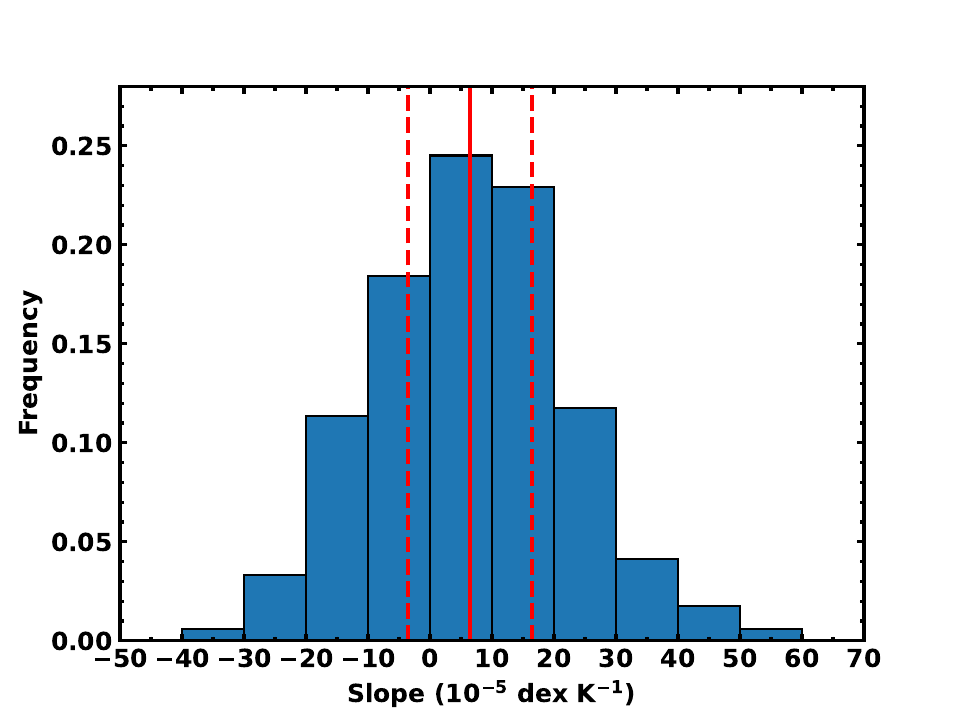}{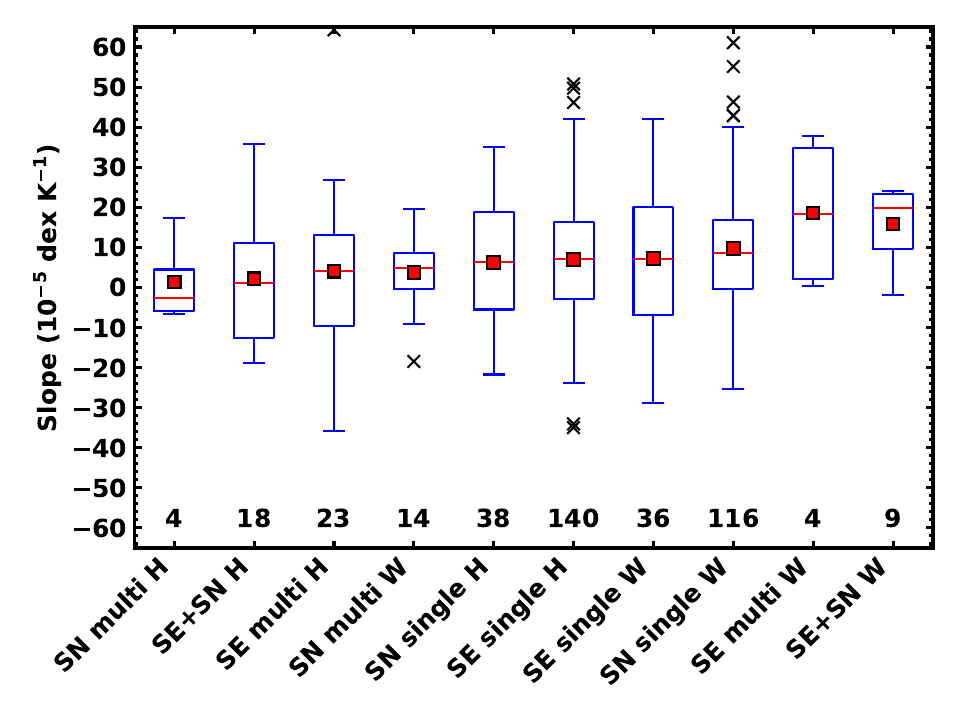}
\caption{\textit{Left panel:} Distributions of slopes of the weighted least-squares (WLS) fits for the chemical abundances [X/Fe] versus the condensation temperatures T$_{C}$. The red solid and dashed lines represent the median and MAD, respectively. \textit{Right panel:} Box plots for the distributions of slopes for different systems, which are organized in order of ascending median slope. The red lines and squares represent the median and mean slopes, respectively. The blue boxes extend from the 25th (Q1) to the 75th (Q3) percentile. The whiskers, which cover the range between Q1-1.5IQR and Q3+1.5IQR, where IQR is the inner quartile range, are shown by blue lines. The outliers are represented by black crosses, and the number of systems in each class can be seen in the lower part of the figure.
\label{fig_slopes}}
\end{figure*}

The distribution of slopes obtained for our sample is shown in the left panel of Figure \ref{fig_slopes}. There are only 11 stars in our sample for which both the Pearson and Spearman tests returned a $p$-value $<$ 0.001, which means that the correlations are statistically significant. The median slope (6 $\pm$ 10) $\times 10^{-5}$ dex K$^{-1}$ is slightly positive but still compatible with zero within the MAD. Approximately 66\% of our stars have positive slopes, which means that the Sun is refractory depleted relative to most of the stars in our sample.

We also analyzed if there were correlations between slopes and the different types of planetary systems using the Cucconi nonparametric test to evaluate possibly significant differences ($p$-value $<$ 0.001) between the distributions. We computed median slopes for the same pairs of subsamples listed in Table \ref{table_stat_tests}. Although all of these tests revealed no significant differences between the subsamples considered, we notice that in general systems with only hot planets have smaller median slopes than systems with only warm planets. This can be seen in the right panel of Figure \ref{fig_slopes}, which shows the distributions of slopes for each subsample and is sorted in order of ascending median slope (red horizontal lines). With the exception of SN multi W systems, the left part of the figure (lower median slopes) is populated with stars that host only hot exoplanets.

\begin{figure*}
\epsscale{1.0}
\plotone{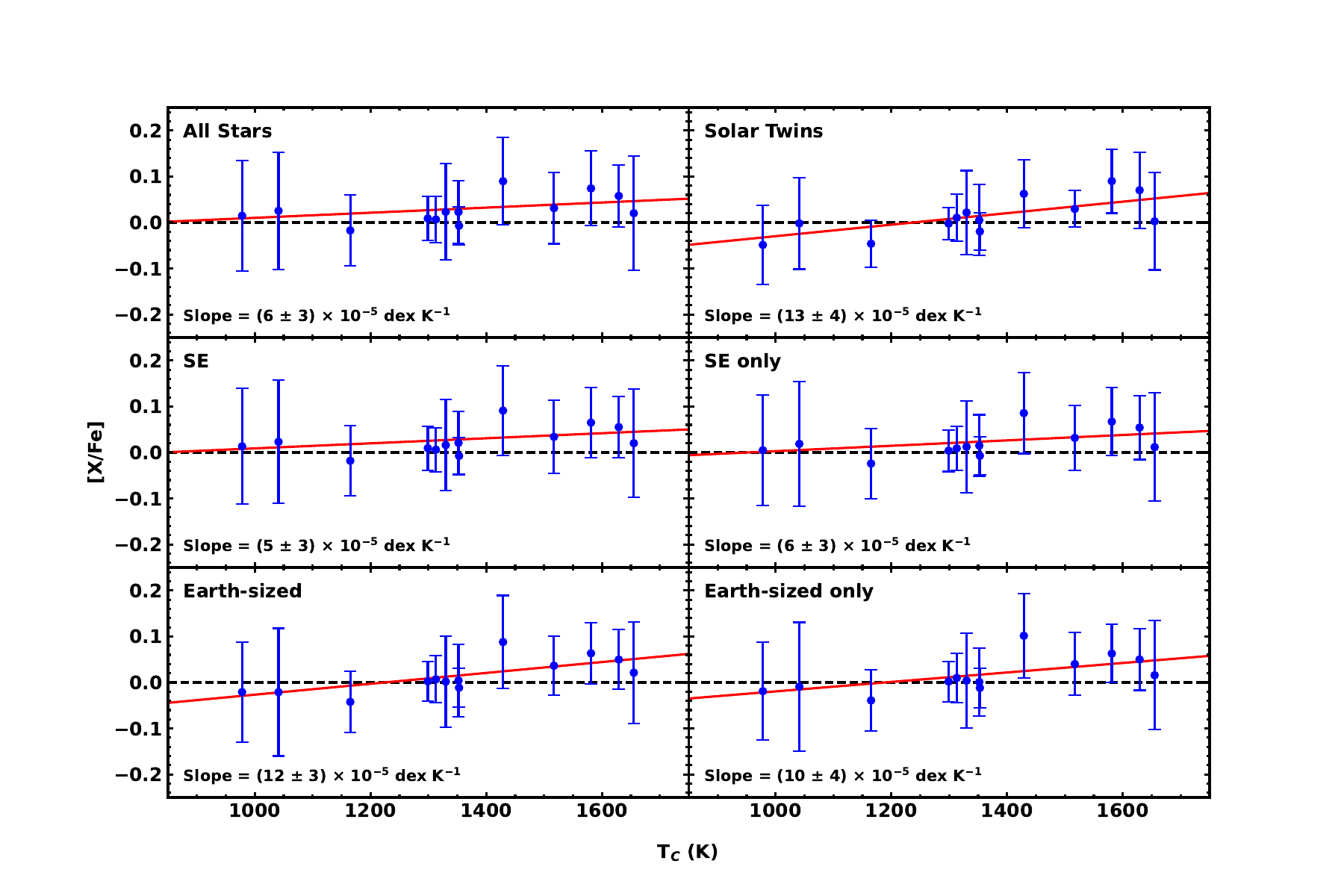}
\caption{Average abundances and standard deviations for different samples: all stars (upper left panel), solar twins (upper right panel), systems with super-Earths (SE; middle left panel), systems with super-Earths only (SE only; middle right panel), systems with Earth-sized planets (lower left panel), and systems with Earth-sized planets only (lower right panel). The red solid lines represent the weighted least-squares fits. The corresponding slopes are provided in each panel. The black-dashed lines show the solar abundance ([X/Fe] = 0 by definition) as a reference.
\label{ab_slopes_twins_se_earths}}
\end{figure*}

As an additional test, we determined the average [X/Fe] abundances for our entire sample of 510 stars and performed a WLS fit as a function of the condensation temperatures of the elements. We show the fit in the upper left panel of Figure \ref{ab_slopes_twins_se_earths} and the calculated slope is (6 $\pm$ 3) $\times 10^{-5}$ dex K$^{-1}$, with $t$-value = 1.630 and $p$-value = 0.131. Thus, it is consistent with zero, and the Spearman and Pearson tests show that the correlation is not significant. 

As previously discussed in the literature \citep[e.g.,][]{schuler15,bedell18,liu20,rampalli25}, the trends between chemical abundances and condensation temperatures of elements might be affected by Galactic Chemical Evolution (GCE). In this context, we tested correcting our [X/Fe] abundances for the correlations with ages using linear relations and the coefficients in Table \ref{table_gce} (but see the caveats of this assumption as discussed by \citealt{rampalli25}). We assume that the corrected abundances are given by [X/Fe]$_{corr}$ = [X/Fe] - (a$_{X}$ $\times$ Age) + b$_{X}$, where X is a given element. Using [X/Fe]$_{corr}$, we repeated the above tests and found some interesting differences. First, there is a displacement of the slope distribution toward more negative values. The median slope is (0 $\pm$ 10) $\times 10^{-5}$ dex K$^{-1}$, and $\sim$50\% of our stars are more refractory depleted relative to the Sun (i.e., have negative slopes), thus placing it as a typical star (see also \citealt{rampalli25}). Moreover, the slope for the average abundances of the entire sample decreases to (0 $\pm$ 1) $\times 10^{-5}$ dex K$^{-1}$ and it is still statistically consistent with zero. Therefore, these results reinforce that the Sun has a similar or lower content of refractory elements relative to our sample of planet-hosting stars.

\cite{melendez09} found a break in the slopes of the abundances versus condensation temperatures for the Sun compared to those of the solar twins at $T_{C} \approx $ 1200 K, and this limit was also used by \cite{liu20} to separate volatile and refractory elements. If, for example, we simply remove the elements Na, Mn and Cu, which have $T_{C} < $ 1200 K, and recompute the slopes for our sample, for the [X/Fe] abundances, the median slope increases to (10 $\pm$ 17) $\times 10^{-5}$ dex K$^{-1}$, and there is a larger percentage of stars with positive slopes (74\%). However, we still do not find significant differences between the slope distributions for the different classes of exoplanetary systems. For [X/Fe]$_{corr}$, there are negligible changes in the median slope and the percentage of positive slopes. 

Since the refractory depletion in the Sun has been proposed in the literature to be caused by the formation of terrestrial planets \citep{melendez09}, we also investigated the slopes obtained when using the average [X/Fe] abundances for systems with at least one super-Earth (N = 291; left middle panel of Figure \ref{ab_slopes_twins_se_earths}) or super-Earths only (N = 221; right middle panel of Figure \ref{ab_slopes_twins_se_earths}). We determined the values (5 $\pm$ 3) $\times 10^{-5}$ dex K$^{-1}$ for host stars having at least one super-Earth and (6 $\pm$ 3) $\times 10^{-5}$ dex K$^{-1}$ for host stars having super-Earths only, both of which have corresponding $p$-values $> 0.001$. The Pearson and Spearman tests also show that the correlations are not significant. We note that the slopes are also consistent with zero when the abundances corrected for age trends are used.

We also computed the slopes of systems with at least one (N = 42) or only (N = 35) Earth-sized planets ($0.9 R_{\oplus} \leq R_{pl} \leq 1.1 R_{\oplus}$) and found positive values for both: (12 $\pm$ 3) $\times 10^{-5}$ dex K$^{-1}$ and (10 $\pm$ 4) $\times 10^{-5}$ dex K$^{-1}$, respectively. However, the $p$-values for the slopes and the Pearson and Spearman tests are equal to or larger than 0.001. Using [X/Fe]$_{corr}$, we find slopes different from zero within uncertainties, and the associated $p$-value is lower than 0.001 only for the systems with at least one Earth-sized planet. However, the Pearson and Spearman tests return $p$-values $>$ 0.001 for both cases. In summary, these comparisons suggest that the Sun has, at most, the same amount of refractories relative to stars hosting Earth-sized planets as well as other planets (which are all sub-Neptunes for this subsample) and stars hosting only Earth-sized planets. Since the Sun is typically more depleted in refractories even when compared with stars that host Earth-sized planets, this suggests its chemical peculiarity might be explained by factors other than planet formation.

\subsubsection{Solar twins}
\label{sec:slopes_twins}

Focusing now on host stars with parameters very similar to the solar values, previous works in the literature have found that the Sun is depleted in refractory elements relative to solar twins \citep[e.g.,][]{melendez09,bedell18,rampalli24,sun25a,sun25b}. If we assume here that solar twins are stars that have stellar parameters within the ranges 5677 K $\leq$ \teff\ $\leq$ 5877 K, 4.34 $\leq$ \logg\ $\leq$ 4.54 and -0.10 dex $\leq$ \feh\ $\leq$ 0.10 dex, there are 25 solar twins in our sample. Among these systems, we have 6 SE, 11 SN, 6 SE + SN, 1 SN + SS, and 1 JP. 

It is worth noting that four of these stars (KOI-444, KOI-3232, KOI-4383, and KOI-4400) have ages within 1 Gyr, masses within 5\%, and radii within 10\% of the solar values. Their average [X/H] abundances are also consistent with the solar pattern within 2$\sigma$ uncertainties: -0.063 $\pm$ 0.037 dex (KOI-444), 0.000 $\pm$ 0.057 dex (KOI-3232), 0.036 $\pm$ 0.089 dex (KOI-4383), and 0.050 $\pm$ 0.069 dex (KOI-4400). 

We estimated their chromospheric activity levels through the log(R$^{'}_{HK}$) index calculated following the methodology described in \cite{lt25}. The derived values are -4.937 (KOI-444), -4.646 (KOI-3232), -4.607 (KOI-4383) and -4.851 (KOI-4400). Thus, KOI-444 and KOI-4400 have chromospheric activity levels that are compatible with the solar values (e.g., -4.937 from \citealt{noyes84}; -4.906 from \citealt{mh08}; and -5.021 from \citealt{lorenzo-oliveira18}). They are also consistent with the values determined by \cite{isaacson24} (-4.981 for KOI-444 and -5.131 for KOI-4400) and place these twins in the regime of inactive stars, log(R$^{'}_{HK}$) $<$ -4.75, as defined by \cite{henry96}. All the parameters discussed above make KOI-444 and KOI-4400 the two best solar twins in our sample. The former hosts a warm sub-Neptune, while the latter hosts a hot Earth-sized planet. To the best of our knowledge, this is the first time that these two solar twins are identified in the literature with such a detailed characterization.

To evaluate the slopes for the subsample of solar twins, we performed a WLS fit to the average abundances of all 25 stars discussed above (using the standard deviations as the weights). As before, we use only [X/Fe] abundances to minimize the effects of atomic diffusion \citep[e.g.,][]{dotter17,bedell18}. We found a positive slope of (13 $\pm$ 4) $\times 10^{-5}$ dex K$^{-1}$, with $t$-value = 3.165 and $p$-value = 0.009. The fit obtained for this sample is shown in the upper right panel of Figure \ref{ab_slopes_twins_se_earths}, and the Spearman and Pearson tests both confirm that the correlation is not significant ($p$-value $>$ 0.001). The null or positive slope suggests that the Sun is similar or refractory depleted relative to our sample of solar twins, which all have planets in this case. In fact, only 1 of the 25 twins has a negative slope.

If we use the abundances corrected for age trends instead, the correlation is still not significant according to the Spearman and Pearson tests, although the slope is (5 $\pm$ 2) $\times 10^{-5}$ dex K$^{-1}$, which is lower than before but still different from zero within the uncertainty. Another interesting difference is that now 18 out of the 25 solar twins in our sample (72\%) have positive slopes, and consequently, the Sun is more depleted in refractory elements relative to them, a value that is consistent with the range of previous estimates from the literature: $>$80\% from \cite{bedell18}, 70\% from \cite{nibauer21}, 87\% from \cite{rampalli24} and 79 - 95\% from \cite{rampalli25}. One main difference between these previous studies and ours is that all solar twins in our sample host exoplanets, making it unlikely that the chemical signature found in previous studies for the Sun may be caused by planet formation. 

As discussed above, there is one star among the solar twins that hosts only a single Earth-sized planet (KOI-4400), and its slope is (29 $\pm$ 6) $\times 10^{-5}$ dex K$^{-1}$, with $t$-value = 5.069 and $p$-value = 0.0004. The correlation is significant according only to the Pearson test. Using the [X/Fe] abundances corrected for age trends instead, both tests return $p$-value $>$ 0.001. Based on its slope of elemental abundances versus condensation temperature, this star is similar or significantly more enhanced in refractories relative to the Sun, although it hosts an Earth-sized planet.

Since it seems that the slope in the case of solar twins is mainly caused by the three elements with $T_{C} < 1200$ K (Na, Mn, and Cu; see Figure \ref{ab_slopes_twins_se_earths}), we removed them from the computation of the slopes determined from the average abundances and repeated the tests above. Now, the slope is consistent with zero for the solar twins as well as  for KOI-4400 and also for the other five subsamples shown in Figure \ref{ab_slopes_twins_se_earths}. In summary, all our results support the conclusion that the depletion of refractory elements in the Sun is not caused by the presence of planets and/or the system's architecture.

\section{Conclusions} 
\label{sec:conclusions}

We determined chemical abundances for 13 elements and 16 species (Na, Mg, Al, Si, Ca, Sc I, Sc II, Ti I, Ti II, V, Cr I, Cr II, Mn, Co, Ni, and Cu) for a sample of 561 stars with planets detected by the Kepler mission \citep{borucki10}. We observed that the abundances for all elements are higher than the solar values (with a median of 0.06 dex), possibly reflecting the higher metallicities of the stars. Although the typical S/N of the spectra is $\sim$60, we obtained overall precise abundances, and the largest median uncertainty is 0.053 for Cr II. 

We analyzed the abundances [X/H] as a function of planetary radii as well as orbital period and observed that stars hosting large planets are statistically enriched for most elements relative to those hosting small planets. A similar result was not obtained, though, when we used [X/Fe] abundances, revealing that the elements are simply following the underlying behavior observed for Fe in \cite{ghezzi21}. 

Going from individual planets to planetary systems, we found statistically significant differences for Al, Si, Sc, Ti, Cr, Fe, and Co when comparing the [X/H] abundances for stars with only small and only large planets. However, this result was not confirmed when we adopted [X/Fe] abundances (corrected or not for trends with age) instead, which means that it was probably caused by the well-known planet-metallicity correlation for giant planets \citep[e.g.,][]{gonzalez97,fischer05,ghezzi18}.

For Al, Si, Ca, Fe, Co, and Ni, we also found that the abundances [X/H] for systems with only hot planets are statistically larger than those of systems with only warm planets, a result that was previously found for iron and a few other elements in the literature \citep[e.g.,][]{mulders16,petigura18,wilson18,wilson22}. Once again, we did not confirm this result with [X/Fe] abundances, suggesting that the other elements simply reflect the higher iron content.

Focusing on systems with only small planets, there are statistically significant differences between Small Hot and Small Warm systems for the abundances [X/H] of Si, Fe, and Co. None of the other multiple comparisons we performed yielded statistically significant results. The analysis of groups of elements yielded consistent, significant differences between systems with only small and only large planets for the refractories, lithophiles and siderophiles. For the comparisons involving [$\alpha$/Fe] and the elemental ratios Fe/Si and Mg/Si, we found no significant differences.

We analyzed the [Fe/H] distributions for low- and high-[$\alpha$/Fe] stars hosting large planets and found that the minimum metallicity is higher for the former. This result suggests that the $\alpha$-elements are able to partially compensate for lower metallicities to allow the formation of giant planets \citep[e.g.,][]{adibekyan12a,adibekyan12b}. However, the iron abundance still limits this process since the minimum metallicity of the high-[$\alpha$/Fe] systems with small planets is much lower than their counterparts with large planets. 

We investigated the slopes of the relation between [X/Fe] abundances and the condensation temperatures of the elements and observed that our stars exhibit a variety of slopes, in agreement with previous studies in the literature \citep[e.g.,][]{rampalli24}. Multiple tests did not yield statistically significant differences between the distributions of slopes for distinct planetary systems. To the best of our knowledge, this is the largest sample of planet-hosting stars for which the study of slopes was performed. 

The analysis of 25 solar twins in our sample confirmed that the Sun is refractory depleted relative to 72\% of them, a percentage that is consistent with previous works \citep[e.g.,][]{bedell18,nibauer21,rampalli24,rampalli25}. We also showed that the Sun has at least the same amount of refractory elements or is more depleted than stars hosting super-Earths or Earth-sized planets.

We have identified the stars KOI-444 and KOI-4400 as the best solar twins in our sample since all their atmospheric and evolutionary parameters, abundances, and activity levels are consistent with those of the Sun within the typical uncertainties. To the best of our knowledge, this is the first time that both of them are presented with such a thorough characterization. The solar twin KOI-4400 hosts a single Earth-sized planet and is more depleted in refractory elements or has an abundance pattern similar to the Sun.

In summary, our results reveal that all possible chemical signatures found for the planet-hosting stars are caused by the underlying behavior of iron. Moreover, we did not confirm the hypothesis that the depletion of refractory elements in the Sun could be caused by the formation of terrestrial or giant planets. Although we used the largest sample of planet-hosting stars to date in such studies, the spectra have lower S/N values in general. Therefore, we highlight the importance of continuing the ongoing efforts to obtain high-quality spectra and determine more precise physical parameters and chemical abundances for stars with planetary systems in order to fully understand their formation, properties, and architectures. 


\begin{acknowledgments}
We warmly thank the California-Kepler Survey team for making
their Keck HIRES reduced data available. We thank the referee and the statistics and data editors for suggestions that helped improve the paper. V.L.T. acknowledges support from the CNPq through the Postdoctoral Junior (PDJ) fellowship, process No. 152242/2024-4. This work has made use of the SIMBAD database, operated at CDS, Strasbourg, France; the VALD database, operated at Uppsala University, the Institute of Astronomy RAS in Moscow, and the University of Vienna; and NASA’s Astrophysics Data System (ADS) and Exoplanet Archive.
\end{acknowledgments}




%
\facilities{Keck:I (HIRES)}

\software{ARES \citep{sousa15}, 
         {Astropy \citep{astropy:2013, astropy:2018}}, 
         {IRAF \citep{tody86,tody93}}, 
         {MOOG \citep{sneden73}}, 
         {matplotlib \citep{Hunter:2007}}, 
         {numpy \citep{2020NumPy-Array}}, 
         {pandas \citep{mckinney-proc-scipy-2010}}, 
         {scipy \citep{2020SciPy-NMeth}}
         }



\bibliography{sample701}{}
\bibliographystyle{aasjournalv7}



\end{document}